\documentclass[11pt]{article}
\pdfoutput=1

\usepackage[T1]{fontenc}
\usepackage[utf8]{inputenc}
\usepackage[english]{babel}

\usepackage{setspace}
\usepackage{simplewick}
\usepackage[dvipsnames]{xcolor}
\usepackage[normalem]{ulem}
\usepackage{cite}
\usepackage{dsfont}
\usepackage{marginnote}
\usepackage{placeins}
\usepackage{xcolor}
\usepackage{caption}
\usepackage{subcaption}

\usepackage{tikz}
\usepackage{pgfplots}
\pgfplotsset{compat=1.18}
\usetikzlibrary{decorations,calligraphy}
\usetikzlibrary{decorations.pathmorphing}
\tikzset{snake it/.style={decorate, decoration=snake}}
\usetikzlibrary{calc}
\usetikzlibrary{math}
\usepackage{float}

\def\centerarc[#1](#2)(#3:#4:#5)
{ \draw[#1] ($(#2)+({#5*cos(#3)},{#5*sin(#3)})$) arc (#3:#4:#5); }

\usetikzlibrary{intersections}
\usetikzlibrary{patterns, arrows.meta, scopes, fadings, intersections, pgfplots.fillbetween}
\usepackage{blindtext}
\usepackage{feynmf}

\newcommand{\de}{\text{d}}

\newcommand{\xdownarrow}[1]{
  {\left\downarrow\vbox to #1{}\right.\kern-\nulldelimiterspace}
}

\usepackage{geometry}
\usepackage{verbatim}
\usepackage{prettyref}
\usepackage{amsmath}
\usepackage[normalem]{ulem}
\usepackage{amssymb}
\usepackage{graphicx}
\usepackage{esint}
\usepackage{verbatim}
\usepackage{physics}
\usepackage{listings}
\usepackage{dsfont}

\usepackage{bm}
\usepackage{mathtools}
\usepackage{bbm}

\usepackage{color}
\definecolor{darkgreen}{rgb}{0,0.5,0}
\definecolor{darkblue}{rgb}{0,0,0.6}
\definecolor{purple}{rgb}{0.4,.2,0.7}
\definecolor{orange}{rgb}{0.95, 0.5, 0.3}
\definecolor{gravity}{rgb}{0.88, 0.92, 1}

\usepackage[colorlinks=true,citecolor=darkblue,linkcolor=darkblue,urlcolor=blue]{hyperref}

\numberwithin{equation}{section}
\numberwithin{table}{section}

\def\cD{\mathcal D}

\def\cF{\mathcal F}

\def\cM{\mathcal M}

\def\cO{\mathcal O}

\def\be{\begin{equation}}
\def\ee{\end{equation}}
\def\bea{\begin{eqnarray}}
\def\eea{\end{eqnarray}}
\def\ba{\begin{align}}
\def\ea{\end{align}}

\def\cO{{\cal O}}
\def\de{\text{d}}

\definecolor{codegreen}{rgb}{0,0.6,0}
\definecolor{codegray}{rgb}{0.5,0.5,0.5}
\definecolor{codepurple}{rgb}{0.58,0,0.82}
\definecolor{backcolour}{rgb}{0.95,0.95,0.92}

\lstdefinestyle{mystyle}{
    backgroundcolor=\color{backcolour},   
    commentstyle=\color{codegreen},
    keywordstyle=\color{magenta},
    numberstyle=\tiny\color{codegray},
    stringstyle=\color{codepurple},
    basicstyle=\ttfamily\footnotesize,
    breakatwhitespace=false,         
    breaklines=true,                 
    captionpos=b,                    
    keepspaces=true,                 
    numbers=left,                    
    numbersep=5pt,                  
    showspaces=false,                
    showstringspaces=false,
    showtabs=false,                  
    tabsize=2
}

\makeatletter
\newcommand{\vast}{\bBigg@{4}}
\newcommand{\Vast}{\bBigg@{5}}
\makeatother

\begin{document}
\begin{spacing}{1.2}
  \setlength{\fboxsep}{3.5\fboxsep}

~
\vskip5mm

\begin{center} 

{\Huge \textsc \textbf{Gravity as a mesoscopic system}}

\vskip10mm

\thispagestyle{empty}

Pietro Pelliconi$^{1,2}$, Julian Sonner$^{1}$ and Herman Verlinde$^{2}$ \\
\vskip5mm
{\small
{\it 1) Department of Theoretical Physics, University of Geneva \\ 24 quai Ernest-Ansermet, 1211 Genève 4, Suisse}
\vskip5mm

{\it 2) Department of Physics, Princeton University, Princeton, NJ 08544}}
\vskip9mm

{ \hypersetup{allcolors=black}{ \url{pietro.pelliconi@unige.ch} , \url{julian.sonner@unige.ch} , \\
\url{verlinde@princeton.edu} } }

\end{center}

\vskip10mm

\begin{abstract}
\noindent We employ a probabilistic mesoscopic description to draw conceptual and quantitative analogies between Brownian motion and late-time fluctuations of thermal correlation functions in generic chaotic systems respecting ETH. In this framework, thermal correlation functions of `simple' operators are described by stochastic processes, which are able to probe features of the microscopic theory only in a probabilistic sense. We apply this formalism to the case of semiclassical gravity in AdS$_3$, showing that wormhole contributions can be naturally identified as moments of stochastic processes. We also point out a `Matryoshka doll' recursive structure in which information is hidden in higher and higher moments, and which can be naturally justified within the stochastic framework. We then re-interpret the gravitational results from the boundary perspective, promoting the OPE data of the CFT to probability distributions. The outcome of this study shows that semiclassical gravity in AdS can be naturally interpreted as a mesoscopic description of quantum gravity, and a mesoscopic holographic duality can be framed as a moment-vs-probability-distribution duality.

\end{abstract}

\pagebreak
\setcounter{page}{1}
\pagestyle{plain}

\setcounter{tocdepth}{2}
{}
\vfill
\tableofcontents

\section{Introduction}

The concept of universality, the insight that certain classes of observables of physical systems are independent of the microscopic details of the theory, finds a wide variety of application in modern statistical mechanics. The central idea revolves around the fact that considering a system with a large number of constituents, microscopic details of the theory cease to be important, and emergent phenomena depend only on a few globally defined parameters. 

Celebrated examples of this idea range from critical exponents in phase transitions, to Random Matrix Theory (RMT) universality classes of quantum chaotic systems \cite{Zinn-Justin_book, PhysRevLett_52_1, BGS_laplacian, Altland_Zirnbauer_classification}. Moreover, a notion of universality can also be found in a disguised form in gravitational theories. Perhaps the first example of this sort dates back to the {\it no hair} theorem for black holes \cite{No_Hair_1, No_Hair_2, No_Hair_3}, where any possible internal structure of a black hole could be described from the outside by a handful of parameters, such as its rest energy $M$, its spin $J$ and its charge $Q$. The connection with statistical mechanics was then later consolidated in famous works of Bekenstein \cite{Bekenstein:1972tm, Bekenstein:1973ur} and Hawking \cite{Hawking:1974rv, Hawking:1975vcx}, realizing that black holes are true thermal objects, which in the certain string-theory examples can be given a statistical interpretation in terms of microstates \cite{Strominger:1996sh}.

This broad vision proved very successful in recent years, where ideas revolving around the universality of quantum chaotic systems were applied to study quantum properties of black holes, in particular in connections with RMT \cite{Cotler:2016fpe, Saad:2018bqo, Saad:2019lba, Saad:2019pqd, Cotler:2020ugk, Altland:2020ccq, Jafferis:2022uhu, Jafferis:2022wez, DiUbaldo:2023qli} and the Eigenstate Thermalization Hypothesis (ETH) \cite{Sonner:2017hxc,Pollack:2020gfa,Belin:2020hea, Altland:2021rqn, Chandra:2022bqq, Belin:2023efa}, the idea that eigenvectors of chaotic systems are randomly distributed \cite{Mark_Srednicki_1999}. The outcome of these studies underlined the chaotic nature of black holes, and connected the aforementioned black hole universality with quantum chaotic universality. In this work, we explore a new connection between gravitational universality and statistical mechanics, in particular between thermal correlation functions of light operators and stochastic processes. 

The theory of stochastic processes dates back to early studies of motions of particles suspended in fluids \cite{Einstein_Brownian, Langevin_1908}, which proved very successful in the confirmation of the atomic theory \cite{Perrin_1909}, and later inspiring new avenues for mathematics \cite{Ornstein_Uhlenbeck_1930}. The idea lies in promoting the deterministic microscopic description of a system to a probabilistic one. The main virtue of the latter resides in its relative simplicity, yet being able to describe emergent phenomena. The paradigmatic example of this class of descriptions is the Brownian motion of a macroscopic particle suspended in a fluid. The highly chaotic interaction between this particle and the fluid molecules results in a stochastic differential equation of the form
\begin{equation}
    \de_t v(t) = - \, \gamma \, v(t) + \xi(t) \ ,
    \label{SDE}
\end{equation}
where $\xi(t)$ is a stochastic noise. The presence of the fluid affects the particle in two combined ways, a deterministic one via the dynamical viscosity that opposes the motion of the particle, and a stochastic one that takes into account microscopic fluctuations given by the various interaction with the fluid molecules. The resulting mathematical description promotes $v(t)$ to have a probabilistic nature, which is connected with experiments through expectations values. Because of this, $\mathbb E[v(t)]$ slows down exponentially, and eventually vanishes at late times, but $\mathbb E[v^2(t)]$ does not, reaching thermal equilibrium with the surrounding. Such a probabilistic description is sometimes called {\it mesoscopic}, to emphasize that it is positioned in between the impractical microscopic one, and an effective one that loses the emergent phenomena of statistical mechanics. 

Our main aim for this work is to create a connection between the theory of stochastic processes and thermal correlation functions of simple operators in semiclassical gravity in AdS, which we denote by $G_{\beta}(t)$. We expect this connection to hold for generic dimensions, but we will concretely analyze only the case of three dimensional gravity. This will allow us to give analytic results, that can also be reinterpreted from a boundary perspective.

The qualitative picture is similar to the one outlined above for Brownian motion, where the exponential decay at early times becomes subdominant at later times with respect to thermal fluctuations. One is able to describe the exponential decay through the contribution of  {\it trivial topology}, interpreted in this context as $\mathbb E[G_{\beta}(t)]$, while the late-time thermal fluctuations are captured by $\mathbb E[G^2_{\beta}(t)]$, which in gravity is represented by a {\it two-boundary wormhole}. Perhaps incidentally, the crossover between the two regimes happens at the timescale $t \sim S$, which is the same order as the Page time. Even though our framework is rather different than Page' set up, it is interesting to notice that such timescale seems the earliest at which a microscopic structure of quantum gravity emerges in some coarse grained way. Another remarkable outcome of our study is that, in a particular double scaling limit, the gravitational answer is consistent with the Brownian paradigm. 

The resulting picture suggests that semiclassical gravity is a mesoscopic description of quantum gravity, where a few global parameters are able to capture emergent phenomena, such as the size of thermal fluctuations, from the microscopic chaotic theory, but crucially where certain microscopic features, such as the erratic fluctuations around the macroscopic means are also consistently accounted for.

The plan of the paper is as follows. 

In Section~\ref{BM_and_thermal_two_pt} we give more details about the emergence of a mesoscopic description, and we analyze the simple example of Gaussian Brownian motion. We then introduce the class of observables that we will analyze in gravitational systems. It is interesting to observe that this mesoscopic interpretation arises for generic quantum chaotic system, following from ETH.

In Section~\ref{sec:Gravity_expectations}, we then apply these notions to thermal correlation functions of light operators in AdS$_3$, in a semiclassical approximation. We describe in detail how a probabilistic interpretation naturally identifies wormhole geometries as moments of probability distributions, which can then be used to infer features of the microscopic theory. In this sense, semiclassical gravity can be understood as a mesoscopic description of quantum gravity. We also point out a recursive structure for which the fluctuations of a fixed moment can be obtained computing a higher moment. This is a natural outcome in our framework where fluctuations are only accounted for in a probabilistic sense, and highlights how finer and finer information is hidden in higher moments.

In Section~\ref{sec:Mesoscopic_CFTs}, we give a boundary interpretation of the results of Section~\ref{sec:Gravity_expectations}, which is more directly connected with the formulation described in Section~\ref{BM_and_thermal_two_pt}. In particular, the mesoscopic description arises considering the OPE data of the CFT as probability distributions. It is quite intriguing to notice that, while the gravitational mesoscopic description is naturally formulated in terms of moments, the CFT side of the story is more naturally formulated in terms of probability distributions. In the mesoscopic framework then, the duality between the two can be framed as a duality between moments and probability distributions, which interestingly implies already a number of constraints on the moments, by mathematical consistency.

We then conclude with Section~\ref{sec:Discussion} discussing the results obtained and laying down future directions to be explored.

\section{Stochastic processes and quantum chaotic systems} \label{BM_and_thermal_two_pt}

The distinctive aspect of a mesoscopic description lies in its probabilistic nature. This feature, which arises when coarse graining a microscopic theory, is effective only in specific contexts. The main reason is that, from a microscopic point of view, any observable that we want to describe is fundamentally {\it not} probabilistic in nature. For example, a simple interaction between a particle and a fluid molecule is completely deterministic, and in most cases (perhaps approximately) solvable. The core idea of a mesoscopic description relies on the fact that, when considering a many-body problem involving a large number of particles, it exists a time-window for which a stochastic interpretation is valid.  

To be more precise, a mesoscopic description of a system holds whenever the following three conditions are met.
\begin{enumerate}

    \item A thermodynamic limit has to be performed, where the number of particles of the system becomes very large. 

    \item There must be a clear scale separation between a probe and an environment. 

    \item The dynamics has to be observed for a long stretch of time, much longer than the intrinsic fluctuations of the system. In this way such fluctuations become effectively random, and a stochastic interpretation holds. For a particle in a fluid, this timescale is the mean free time, while for a correlation function is the inverse spectral width (in natural units). 

\end{enumerate}
The first condition ensures that the system under consideration presents emergent phenomena. The second condition is necessary in order to have a clear scale separation between the low-energy physics and the microscopic one. The third condition sets the smallest timescale at which the probabilistic interpretation holds. 
All three conditions above are met in holographic CFTs. We can think of the large-$c$ limit as a thermodynamic limit, where a large number of degrees of freedom is considered. Furthermore in this regime, a theory with a bulk-local dual has a clear spectral gap between a handful of light operators, and a dense spectrum of heavy ones, which is the scale separation needed to have a well defined observable in the above sense. Finally, the spectral width is parametrically large in $c$, effectively sending to zero the interaction time.  More details will be presented in Section~\ref{sec:Mesoscopic_CFTs}.

We now briefly review the most celebrated example of a stochastic process, Brownian motion. Quite remarkably, thermal correlation functions in semiclassical gravity will belong to the same universality class.

\subsection{A paradigmatic example} \label{sec:Brownian_motion}

A paradigmatic example for a mesocopic description is the case of a macroscopic particle suspended in a fluid, and its motion thereof\footnote{For textbook references on stochastic methods see \cite{Gardiner_Stochastic_Methods} or Lennart Sjögren's lecture notes on Brownian motion~\cite{Lennart_Lecture_notes}.}. For simplicity, we focus on a particle surrounded by an ideal gas. A good approximation to the laws of motion of the macroscopic particle is 
\begin{equation}
    \de_t v(t)   = - \, \gamma \, v(t) \ ,
    \label{law_of_dynamical_viscosity}
\end{equation}
which describes a viscous force that opposes the free motion of the particle. The solution to~\eqref{law_of_dynamical_viscosity} vanishes exponentially at late time, and fails to describe the thermal fluctuations in the velocity of the particle. This is clearly in contrast with statistical mechanics, which demands that at late times the particle will have the same (inverse) temperature $\beta$ of the fluid, and thus for Maxwell's kinetic theory
\begin{equation}
    \mathbb E[v^2(t)] \, \to \,  \frac{1}{m \beta} \qquad \text{for} \quad t \to \infty \ ,
    \label{Maxwell_mean_velocity_squared}
\end{equation}
where $m$ is the mass of the particle. Perhaps remarkably, a simple modification to~\eqref{law_of_dynamical_viscosity} that {\it does} take into account thermal fluctuations introduces a stochastic noise in the equation of motion, resulting in the Langevin equation
\begin{equation}
    \de_t v(t) = - \gamma \, v(t) + \xi(t) \ .
    \label{Langevin}
\end{equation}
In this framework, the velocity of the particle is interpreted in a probabilistic sense, and only expectation values can be compared with observations. Assuming the stochastic noise to have mean zero and to be uncorrelated, so that $\mathbb E[\xi(t)] = 0 $ while $\mathbb E[\xi(t) \xi(s)] = g \, \delta(t - s)$, it is not hard to prove that on average the particle follows the effective equation~\eqref{law_of_dynamical_viscosity}, since
\begin{equation}
    \de_t \, \mathbb E[v (t)] = - \gamma \, \mathbb E[v (t)] \qquad \text{so that} \qquad \mathbb E[v (t)] = v_0 \, e^{- \gamma t } \ ,
    \label{Brownian_solution_one_pt}
\end{equation}
while higher moments are affected by the stochastic force, 
\begin{equation}
    \mathbb E[ v(t) v(s) ] = \mathbb E[ v(t) ] \, \mathbb E[ v(s) ] - \frac{g}{2 \gamma} \, e^{- \gamma (t + s)} + \frac{g}{2 \gamma} e^{- \gamma |t - s|} \ , \qquad \text{so that} \qquad \mathbb E[ v^2(t) ] \to \frac{g}{2 \gamma} \ .
    \label{Brownian_two_pt}
\end{equation}
The phenomenological parameter $g = 2 \gamma / m \beta$ is then chosen to match the thermal fluctuations~\eqref{Maxwell_mean_velocity_squared} of the system, and being related to the dynamical viscosity $\gamma$ of the system it is an instance of a {\it fluctuation-dissipation} theorem.

In summary, Brownian motion is a mesoscopic description of the dynamics of a particle in a fluid, giving an intermediate description between a macroscopic one and a fully microscopic one. In particular, it allows to capture both the large-scale emergent dynamics seen at a macroscopic scale and the microscopic fluctuations, albeit in a probabilistic way. This latter feature is in contrast to the true microscopic description where the fluctuations are explicitly and not stochastically included, but which often is too fine-grained to be able to effectively capture emergent macroscopic phenomena.

In what follows, we will argue that the semiclassical gravity path integral furnishes exactly such a mesoscopic description of gravity, and we demonstrate that the statistical distribution of fluctuations is determined by multi-boundary `wormhole' type contributions to the path integral. In order to do so, in the next Section we introduce the class of observables of interest to us.

\subsection{Correlation functions of quantum chaotic systems} \label{sec:Correl_funct_quantum_chaotic_systems}

The main players of this work will be thermal correlation functions of such simple operators. In the most general form, this work addresses correlations functions of the form
\begin{equation}
    G_{\beta}(t_1, \tau_1; \dots ; t_n, \tau_n) = \frac{1}{Z(\beta)} \Tr \Big[ \, e^{- \tau_n H} \, \cO(t_n) \, \dots \, e^{- \tau_1 H} \, \cO(t_1)  \Big] \ ,
    \label{generic_thermal_corr_funct}
\end{equation}
where $\tau_1 + \dots + \tau_n = \beta$, the (inverse) temperature of the system. Said differently, we consider correlation functions between operators placed around the thermal circle, and appropriately continued to Lorentzian times. These are a natural class of observables to consider because they define low-energy measurements accessible to low-energy observers. Moreover, these types of observables \eqref{generic_thermal_corr_funct} describe the state of a black hole under evaporation \cite{Stanford:2020wkf}, emitting Hawking pairs described by the operators $\cO$.

In generic quantum chaotic systems, the operators $\cO$ are {\it e.g.} local operators, while in holographic theories they are low-energy fields that do not heavily backreact on the geometry, or equivalently CFT operators with a small conformal dimension, much smaller than the black hole threshold. Among the class of observables~\eqref{generic_thermal_corr_funct}, we concentrate on the simplest one, namely
\begin{equation}
    G_{\beta}(t) =  \frac{1}{Z(\beta)} \Tr \left[ e^{- \beta H/2} \, \cO(t) \, e^{- \beta H/2} \, \cO(0)  \right]  \ ,
    \label{Two_pt_probe}
\end{equation}
a two point correlation function where the operators are placed symmetrically with respect to the thermal circle. Notice that, for simplicity, we have suppressed the variables $\tau_1 = \tau_2 = \beta / 2$ in the argument of the function. We will also assume a finite system size, in order to have a discrete spectrum. By definition, the density of eigenvalues will then be related to the micro-canonical entropy of the energy window around $E$ by the relation $\rho(E) \sim e^{S(E)}$.

Before focusing to gravity, it is instructive to understand if a mesoscopic description is generically applicable to quantum chaotic systems, and what kind of predictions one is able to find. Besides the few necessary assumption spelled out at the beginning of Section~\ref{BM_and_thermal_two_pt}, we consider chaotic system respecting the Eigenstate Thermalization Hypothesis (ETH) \cite{Mark_Srednicki_1999}, which gives a general statement about the matrix elements $\cO_{ij} = \bra{E_i} \cO \ket{E_j}$ of `simple' operators in the microcanonical window $E = ( E_i + E_j)/2$. In particular, ETH assumes that they are randomly distributed\footnote{In the formulation~\eqref{ETH} of ETH we have assumed $\mathbb E[\cO_{ii}] = 0$. If this is not the case, we can just consider the shifted operator $\tilde \cO_{ij} = \cO_{ij} - \mathbb E[\cO_{ii}]$ .}
\begin{equation}
    \cO_{ij} = e^{-S(E)/2} \, F_{1}(E,\omega) \, R_{ij} \ , \qquad \text{such that} \qquad \mathbb E \left[|\cO_{ij}|^2 \right] = e^{-S(E)} \, F_{2}(E, \omega)   \ ,
    \label{ETH}
\end{equation}
where $F_{1,2}(E, \omega)$ are smooth function of the energy $E$ and of the difference $\omega = E_i - E_j$, and $R_{ij}$ are mean-zero random variables. In fact, in recent years, it has been appreciated that the above statement is but the tip of the iceberg of a whole hierarchy of higher moments of the statistical distribution of (connected) moments of the $R_{mn}$ \cite{foini2019eigenstate,Jafferis:2022uhu, Jafferis:2022wez}, which in our notation amounts to
\begin{equation}
\mathbb E \left[\cO_{i_1 j_1} \cO_{i_2 j_2} \dots \cO_{i_n j_n} \right] = e^{-(n-1)S(E)} \, F^{(n)}_{i_1 j_1 \ldots i_n j_n}(E_1, \ldots , E_n)   \ ,
\end{equation}
with $E$ being the average of the $n$ independent energies $E_i$ and the $n$-th moment is encapsulated in the function $F^{(n)}$ analogously to $F_{1,2}$ above\footnote{The index structure of $F^{(n)}_{i_1 j_1 \ldots i_n j_n}$ is such that it can be reduced to a single smooth function of its arguments multiplied by exactly $n$ Kronecker deltas summed over the permutation group $S_n$ \cite{Jafferis:2022uhu, foini2019eigenstate}. }. In essence, ETH describes the fluctuations of matrix elements of simple operators in a probabilistic manner.

With these assumptions, it is possible to draw general conclusions on the class of observables~\eqref{Two_pt_probe}. At early times, the two-point function~\eqref{Two_pt_probe_stochastic} undergoes a phase of decay in $t$. One way to see this proceeds by realising that at early enough times $t \ll e^{S}$, the discreteness of the spectrum is washed out, and we can exchange sums for integrals as
\begin{equation}
    \sum_{n,m} \quad \rightarrow \quad \int \de E_1 \de E_2 \, \rho(E_1) \, \rho(E_2) \approx \int \de E \, \de \omega \, \rho^2(E)   \ ,
    \label{sums_for_integral}
\end{equation}
where in the RHS we wrote the double integral in terms of the average energy $E$ and of the difference $\omega$, and we approximated\footnote{It is also possible to take into account corrections to this result, as in \cite{Jafferis:2022wez}.} $\rho(E + \frac{\omega}{2})\rho(E - \frac{\omega}{2}) = \rho^2(E) + O(\omega^2)$ at linear order in $\omega$. We can then write~\eqref{Two_pt_probe} in the energy eigenbasis, and use ETH to estimate the dynamics of the two point function, finding
\begin{equation}
    G_\beta (t) \, \approx \, \frac{1}{Z(\beta)} \int_{0}^{\infty} \de E \,   \rho(E) \, e^{- \beta E } \int_{- E/2}^{E/2}  \de \omega \, e^{- i \omega t} \, F_2(E, \omega)  \ ,
    \label{approximation_Gb}
\end{equation}
where we have additionally used that $\rho(E) = e^{S(E)}$. As an additional simplifying assumption, we may assume that the dominant contribution in the $E$ integral of~\eqref{approximation_Gb} comes from a saddle-point approximation that localises the microcanonical energy $E$ to the thermodynamical energy $ E_{\beta}$, and that this result is independent of $F_2(E, \omega)$. Thus, the expression above reduces to 
\begin{equation}
    G_\beta (t) \, \approx \,  \int \de \omega \, e^{- i \omega t} \, F_2(E_{\beta}, \omega)  \ ,
    \label{approximation_Gb_final_FT}
\end{equation}
which is a standard Fourier transform in $\omega$ of a smooth function, and we then conclude that the result decays faster than any polynomial in $t$. This however cannot be, since the correlation function of a system in a compact background can not decay forever, as a consequence of the Riemann-Lebesgue lemma. We thus find ourselves in a similar situation as Section~\ref{sec:Brownian_motion}. Our effective description of the system loses information on the late time fluctuations of the system, while a microscopic description of it is impractical to observe emergent phenomena. 

We are then in a position to explore a mesoscopic description, which is in fact given by ETH itself. The stochastic nature of the correlation function arises from the probabilistic interpretation of the overlaps $\cO_{mn}$, which in turn makes the whole correlation function 
\begin{equation}
    G_{\beta}(t) =  \frac{1}{Z(\beta)} \sum_{i,j} e^{- \frac{\beta(E_i + E_j)}{2}} e^{- i (E_i - E_j) t}   \, \big| \cO_{ij} \big|^2  
    \label{Two_pt_probe_stochastic}
\end{equation}
a stochastic process itself. This description of course loses some microscopic informations on the system, but at the same time recovers universal features {\it in a probabilistic sense}. Moreover, the similarities with Section~\ref{sec:Brownian_motion} go beyond this conceptual interpretation, and it is possible to draw a one-to-one connection between the various contributions. First, we can compute the expectation value
\begin{equation}
    \mathbb E[G_{\beta} (t)] = \int \de \omega \, e^{- i \omega t} \, F_2(E_{\beta}, \omega) \ ,
\end{equation}
which, as for~\eqref{Brownian_solution_one_pt}, decays exponentially fast in time, and loses track of thermal fluctuations of the system. Then, inspired by Section~\ref{sec:Brownian_motion}, we can recast microscopic informations on the late time fluctuations computing the second moment of the correlation function. Without loss of generality, we can assume the following ansatz
\begin{equation}
    \mathbb E \left[ \big| \cO_{ij} \big|^2 \, \big| \cO_{kl} \big|^2 \right] = \mathbb E \left[ \big| \cO_{ij} \big|^2\right] \, \mathbb E \left[\big| \cO_{kl} \big|^2 \right] +e^{-2S(E)}  \big ( \delta_{ik} \delta_{jl} + \delta_{il} \delta_{jk} \big ) H_{4}(E, \omega)  \ ,
    \label{disconnected_square_ansatz}
\end{equation}
The first term in the RHS is the disconnected expectation value, which is the dominant one \cite{Foini-Pappalardi-Kurchan_free_prob}. The structure of $\delta$-functions comes from the coefficients being uncorrelated, and this whole subdominant contribution is weighted by a generic smooth function $H_{4}(E,\omega)$, which depends on the specific model under considerations. We can then use the ansatz~\eqref{disconnected_square_ansatz} to find the autocorrelation function, and a straightforward computation gives
\begin{multline}
    \mathbb E \left[ G_{\beta}(t) G_{\beta}(s) \right] \approx \mathbb E[ G_{\beta}(t)] \, \, \mathbb E [G_{\beta}(s)]  + \frac{1}{2 \beta \, Z^2(\beta)}  \int \de \omega \, e^{-i \omega(t + s)} \, H_{4}(E_\beta, \omega) \\
    + \frac{1}{2 \beta \, Z^2(\beta)}  \int \de \omega \, e^{-i \omega(t - s)} \,  H_{4}(E_\beta, \omega)  \ .
    \label{autocorrelation_G_ETH}
\end{multline}
This result is rather constraining. It implies that, besides the disconnected part, two additional contribution arise. One is a function of $t+s$ and the other is a function of $|t-s|$, but {\it precisely with the same functional form.} The same structure is also realized in the case of Brownian motion~\eqref{Brownian_two_pt} previously considered.

\section{Semiclassical gravity as a mesoscopic description} \label{sec:Gravity_expectations}

We now turn to the main result of this work, namely the connection between stochastic processes and semiclassical gravity. In particular, the main aim of this Section is to argue that semiclassical gravity, encompassing quantized matter fields on a geometric background, is an intermediate mesoscopic description of the microscopic one.

We first define in detail the framework we are interested in. We specialize to gravity in asymptotically Anti-de Sitter (AdS) spaces in three dimensions, where analytic results can be explicitly found. The action of interest to us is
\begin{equation}
    I = \frac{1}{16 \pi G_N} \int \de^{3} x \,  \sqrt{h} \left( R + \frac{2}{L^2} \right) + S_{m}[\phi, h] + S_{\rm ct} \ ,
    \label{EH_action}
\end{equation}
where on the RHS we have the Einstein-Hilbert action, the action for light matter fields, and holographic renormalization counterterms \cite{Henningson:1998gx}. We are interested in compact CFTs in thermal states, so we consider boundaries with topology $\mathbb S^1_{\beta} \times \mathbb S^{1}$, the first factor being the thermal circle. The class of observables~\eqref{generic_thermal_corr_funct} can be computed in semiclassical gravity resorting to the prescription 
\begin{equation}
    G_{\beta}(t_1, \tau_1; \dots ; t_n, \tau_n) = \frac{1}{Z(\beta)} \lim_{r \to \infty} r_1^{\Delta} \dots r_n^{\Delta}\sum_{\cM} Z_\cM \int [\cD \phi] \, \phi(x_1, r_1) \dots \phi(x_n, r_n) \, e^{- S[h_{\cM}, \phi]} \ ,
    \label{semiclassical_approx_gravity}
\end{equation}
where the sum is over all saddles that respect the given boundary conditions, namely $\partial \cM = \mathbb S^1_{\beta} \times \mathbb S^{1}$. Importantly, notice the factor $Z_\cM$ that comes from evaluating the Einstein-Hilbert action on-shell on the manifold $\cM$. 

We will only consider correlators of conical defects in the bulk, whose dual operator in the CFT has scaling dimension $1 \ll \Delta \ll L/G_N \sim c$. In this limit, the action \eqref{EH_action} becomes
\begin{equation}
    I = \frac{1}{16 \pi G_N} \int \de x^{3} \,  \sqrt{h} \left( R + \frac{2}{L^2} \right) + m \int \de l + S_{\rm ct} \ ,
    \label{EH_action_with_defects}
\end{equation}
and correlators are well approximated by the geodesic approximation
\begin{equation}
    G_{\beta}(x, y) = \int [\cD \phi] \, \phi(x) \, \phi(y) \, e^{- S[h_{\cM}, \phi]} = e^{- m \,  l(x , y)} \ ,
\end{equation}
where $l(x , y)$ is the length of the bulk geodesic connecting the boundary points $x$ and $y$.

\subsection{Stochastic interpretation and wormholes} \label{sec:different_moments}

We then take a light scalar primary, and compute the symmetric correlation function~\eqref{Two_pt_probe}. For simplicity, we place both operators at the same spatial position, and using the geodesic approximation one gets \cite{Maldacena:2001kr}
\begin{equation}
    G_{\beta}(t) =  \frac{\pi^{2 \Delta}}{\beta^{2 \Delta}} \, \cosh^{- 2 \Delta} \left( \frac{\pi t}{\beta} \right) + \dots  \ .
    \label{avg_two_pt_gravity}
\end{equation}
We notice that, after a transient of order $\beta$, the two point function decays exponentially, vanishing at late times. It is therefore oblivious to the tail of erratic fluctuations, as was first noticed in \cite{Maldacena:2001kr}. Corrections to this result comprise a sum over geodesics that wrap around the black hole. In Appendix~\ref{app:Sum_over_images} we present a careful treatment of these contributions, which however do not modify substantially the exponential decay.

The situation looks analogous to the discussion presented in Section~\ref{BM_and_thermal_two_pt}. The correlator~\eqref{avg_two_pt_gravity} is missing thermal fluctuations of size $e^{- S}$ at late times. Therefore, given the exponential decay in time of \eqref{avg_two_pt_gravity}, around timescales of order the entropy $S$, the thermal correlator is not consistent anymore with a unitary microscopic interpretation. From this point of view, it seems then appropriate to consider \eqref{avg_two_pt_gravity} only as an expectation value $\mathbb E[G_{\beta}(t)]$ of an underlying stochastic process, and to recover information about the thermal fluctuations computing the autocorrelation function $\mathbb E[G_{\beta}(t) G_{\beta}(s)]$. Such an observable in gravity requires manifolds with disjoint boundaries, one for each correlation functions. In principle, it is possible that more than one geometrical saddle exists with the same boundary conditions, and we thus consider all of them. Saddles that connect different boundaries are normally called {\it wormholes}, and are represented in Figure~\ref{wormholes_3d_gravity}.

\begin{figure}[t]
\centering
    \begin{tikzpicture}[scale=0.7]

        \draw[very thick] (-5,-2.5) ellipse (1.5 and 1);
        \draw[very thick] plot [smooth, tension=1] coordinates { (-5.5, -2.4)  (-5, -2.6)  (-4.5, -2.4)};
        \draw[very thick] plot [smooth, tension=1] coordinates { (-5.3, -2.54)  (-5, -2.39)  (-4.7, -2.54)};

        \draw[red,fill=red] (-4.2, -0.3) circle (2pt) ;
        \draw[red,fill=red] (-5.8, -0.3) circle (2pt) ;
        \draw[red,fill=red] (-4.2, -2.8) circle (2pt) ;
        \draw[red,fill=red] (-5.8, -2.8) circle (2pt) ;

        \draw[color = red, thick] plot [smooth, tension=1] coordinates { (-4.2, -0.3)  (-5, -0.55)  (-5.8, -0.3)};
        
        \draw[color = red, thick] plot [smooth, tension=1] coordinates { (-4.2, -2.8)  (-5, -3.05)  (-5.8, -2.8)};

        \draw[] (-2.5, -1.25) node {\Large $+$};

        \draw[] (-7.5, -1.25) node {\Large $=$};

        \draw[] (-11.2, -1.2) node {\Large $\mathbb E[G_{\beta}(t) G_{\beta}(s)]$};

        \draw[very thick] (-5,0) ellipse (1.5 and 1);
        \draw[very thick] plot [smooth, tension=1] coordinates { (-5.5, 0.1)  (-5, -0.1)  (-4.5, 0.1)};
        \draw[very thick] plot [smooth, tension=1] coordinates { (-5.3, -0.04)  (-5, 0.11)  (-4.7, -0.04)};


    \begin{scope}[xshift = -5cm]

        \begin{scope}[xshift=10cm]
        \draw[very thick] (0,-2.5) ellipse (1.5 and 1);
        \draw[very thick] plot [smooth, tension=1] coordinates { (-0.5, -2.4)  (0, -2.6)  (0.5, -2.4)};
        \draw[very thick] plot [smooth, tension=1] coordinates { (-0.3, -2.54)  (0, -2.39)  (0.3, -2.54)};

        \draw[red,fill=red] (0.8, -0.3) circle (2pt) ;
        \draw[red,fill=red] (-0.8, -0.3) circle (2pt) ;
        \draw[red,fill=red] (0.8, -2.8) circle (2pt) ;
        \draw[red,fill=red] (-0.8, -2.8) circle (2pt) ;

        \draw[color = red, thick] plot [smooth, tension=1] coordinates { (0.8, -0.3)  (0.7, -1.55)  (0.8, -2.8)};
        
        \draw[color = red, thick] plot [smooth, tension=1] coordinates { (-0.8, -0.3)  (-0.7, -1.55)  (-0.8, -2.8)};

        \fill[color = Periwinkle, opacity = 0.2] plot [smooth, tension=1] coordinates { (-1.5, 0)  (-1.35, -1.25)  (-1.5, -2.5)} -- (-1.5,-2.5) arc (180:360:1.5 and 1) -- plot [smooth, tension=1] coordinates { (1.5, -2.5) (1.35, -1.25) (1.5, 0) } -- (1.5,0) arc (0:-180:1.5 and 1) -- cycle;

        \draw[color = Periwinkle, thick] plot [smooth, tension=1] coordinates { (-1.5, 0)  (-1.35, -1.25)  (-1.5, -2.5)};
        \draw[color = Periwinkle, thick] plot [smooth, tension=1] coordinates { (1.5, 0)  (1.35, -1.25)  (1.5, -2.5)};

        \draw[very thick] (0,0) ellipse (1.5 and 1);
        \draw[very thick] plot [smooth, tension=1] coordinates { (-0.5, 0.1)  (0, -0.1)  (0.5, 0.1)};
        \draw[very thick] plot [smooth, tension=1] coordinates { (-0.3, -0.04)  (0, 0.11)  (0.3, -0.04)};
        \end{scope}

        \draw[] (2.5, -1.25) node {\Large $+$};

        \draw[] (7.5, -1.25) node {\Large $+$};

        \draw[] (12.5, -1.25) node {\Large $\dots$};

        \begin{scope}[xshift=5cm]
        \draw[very thick] (0,-2.5) ellipse (1.5 and 1);
        \draw[very thick] plot [smooth, tension=1] coordinates { (-0.5, -2.4)  (0, -2.6)  (0.5, -2.4)};
        \draw[very thick] plot [smooth, tension=1] coordinates { (-0.3, -2.54)  (0, -2.39)  (0.3, -2.54)};

        \draw[red,fill=red] (0.8, -0.3) circle (2pt) ;
        \draw[red,fill=red] (-0.8, -0.3) circle (2pt) ;
        \draw[red,fill=red] (0.8, -2.8) circle (2pt) ;
        \draw[red,fill=red] (-0.8, -2.8) circle (2pt) ;

        \draw[color = red, thick] plot [smooth, tension=1] coordinates { (0.8, -0.3)  (0, -2.15)  (-0.8, -2.8)};
        
        \draw[color = red, thick] plot [smooth, tension=1] coordinates { (-0.8, -0.3)  (0, -0.95)  (0.8, -2.8)};

        \fill[color = Periwinkle, opacity = 0.2] plot [smooth, tension=1] coordinates { (-1.5, 0)  (-1.35, -1.25)  (-1.5, -2.5)} -- (-1.5,-2.5) arc (180:360:1.5 and 1) -- plot [smooth, tension=1] coordinates { (1.5, -2.5) (1.35, -1.25) (1.5, 0) } -- (1.5,0) arc (0:-180:1.5 and 1) -- cycle;

        \draw[color = Periwinkle, thick] plot [smooth, tension=1] coordinates { (-1.5, 0)  (-1.35, -1.25)  (-1.5, -2.5)};
        \draw[color = Periwinkle, thick] plot [smooth, tension=1] coordinates { (1.5, 0)  (1.35, -1.25)  (1.5, -2.5)};

        \draw[very thick] (0,0) ellipse (1.5 and 1);
        \draw[very thick] plot [smooth, tension=1] coordinates { (-0.5, 0.1)  (0, -0.1)  (0.5, 0.1)};
        \draw[very thick] plot [smooth, tension=1] coordinates { (-0.3, -0.04)  (0, 0.11)  (0.3, -0.04)};
        \end{scope}
        
    \end{scope}

    \draw[] (-5, -4);
        
    \end{tikzpicture}
    \caption{The semiclassical gravity calculation of the autocorrelation function of $G_{\beta}(t)$. Red solid lines show the various contractions of the operators. All these contributions are vanishing in the limit of large $t$ and $s$, except the third one, which depends on $|t - s|$. }
    \label{wormholes_3d_gravity}
\end{figure}

At this point, the precise computation of the autocorrelation function is not completely straightforward. A general answer to this problem is described in \cite{Chandra:2022bqq} in terms of Liouville correlators\footnote{See also \cite{Collier:2023fwi, Collier:2024mgv} for other recent advancements on three-dimensional gravity.}, which in our specific case we write in \eqref{two_pt_fn_thermal}. However, the explicit saddlepoint involves finding a two-boundary wormhole with different complex structures, since we would like to insert the defects at different Lorentzian times. A simplified treatment can be obtained inserting the defects at the same Lorentzian time, and then use geodesics on such manifold as a good approximation to the ones computed in the exact case. This was also the approach undertaken in \cite{Stanford:2020wkf}. In this case, the saddlepoint can be written as 
\begin{equation}
    \de s^2 = \de \rho^2 + \cosh^2(\rho) \, \de \Sigma^2 \ ,
    \label{MM_wormhole}
\end{equation}
as it was shown in \cite{Maldacena:2004rf}. The metric \eqref{MM_wormhole} is a saddlepoint of the Einstein equations in AdS$_3$ provided that $\de \Sigma^2$ is a hyperbolic metric. However, not all hyperbolic metrics represent two-boundary wormholes. A simple example is the hyperbolic plane $\mathbb H_2$, which is the complex Upper Half Plane (UHP) with metric
\begin{equation}
    \de s^2_{\mathbb H_2} = - \frac{4 \, \de z \,  \de \bar z}{(z - \bar z)^2} \ .
\end{equation}
It is not hard to show that this is just a different parametrization of the Poincaré patch of AdS$_3$, see for example \cite{Anous:2022wqh}. In order to have a proper two-boundary wormhole, we need to quotient $\mathbb H_2$ with a Fuchsian groups, which is a discrete subgroups of ${\rm PSL}(2, \mathbb R)$. The choice of ${\rm PSL}(2, \mathbb R)$ is easily explained, since it maps the UHP into itself through
\begin{equation}
    \gamma z = \frac{a z +b}{c z + d} \ , \qquad \text{with} \qquad \gamma = \begin{pmatrix}
        a & b \\
        c & d 
    \end{pmatrix} \in {\rm PSL}(2, \mathbb R) \ .
\end{equation}
Finally, one can take $\Sigma$ to be the fundamental domain of $\mathbb H_2/\Gamma$. In our case, we would like $\Sigma$ to be a torus with two defects. The associated Fuchsian group will then have an hyperbolic element, that sets the topology of the torus, and two elliptic ones, $\gamma_{1,2}$. To match the conical defect with the scaling dimension of the operator, we require \cite{Chandra:2022bqq}
\begin{equation}
    \Tr(\gamma_{1,2}) = 2 \cos(\varphi) \ , \qquad \text{with} \qquad \varphi = \pi \sqrt{1 - \frac{12 \Delta}{c}} \ .
\end{equation}
At this point, we might be worried that a Fuchsian group $\Gamma$ with such properties does not exist. The condition that has to be satisfied in order to have a wormhole between Riemann surfaces with genus $g$ and conical defects is \cite{Chandra:2022bqq}
\begin{equation}
    2 g + \sum_i \left( 1 - \sqrt{1- \frac{12 \Delta_i}{c}} \, \right) > 2 \ .
    \label{condition_hyperbolicity_genus_matter_ops}
\end{equation}
Therefore, for the case of a torus (which has $g = 1$), we see that any matter insertion stabilizes the saddle. On the other hand, the torus without conical defects is not a saddle, as it is known. 

The upshot of the above discussion has been to give a recipe to find such two boundary wormholes. However, an explicit presentation of the group $\Gamma$ and of the metric \eqref{MM_wormhole} will be both cumbersome and not illuminating. Instead, we extract a good approximation at late times. We first present the result, and then we explain how to obtain it. In the limit $t,s \to \infty$ with $(t-s) \sim O(1)$, the leading approximation is
\begin{multline}
    \mathbb E[G_{\beta}(t) G_{\beta}(s)] \approx \mathbb E[G_{\beta}(t)] \, \mathbb E[ G_{\beta}(s)] + \frac{1}{Z^2(\beta)} \frac{\pi^{4 \Delta}}{\beta^{4 \Delta}}  \cosh^{- 2 \Delta} \left( \frac{\pi t}{\beta} \right) \cosh^{- 2 \Delta} \left( \frac{\pi s}{\beta} \right) \\
    + \frac{1}{Z^2(\beta)} \frac{\pi^{4 \Delta}}{\beta^{4 \Delta}}  \cosh^{- 2 \Delta} \left( \frac{\pi (t - s)}{\beta} \right) + \dots \ .
    \label{2D_gravity_expectations}
\end{multline}
To explain in detail the RHS, it is helpful to look at Figure~\ref{wormholes_3d_gravity}. The first contribution is the disconnected one, which comes as the product of two correlations functions. This is the first diagram in Figure~\ref{wormholes_3d_gravity}. The second contribution is a particular Wick contraction where the Lorentzian-evolved operator on one side is contracted with the non-evolved operator on the other side. In the late-time limit each contraction is indistinguishable from $\mathbb E[G_{\beta}(t)]$, which is what we have used. Moreover, it differs from the previous contribution by $1/Z^2(\beta)$, because the torus wormhole action (stabilized by matter defects) vanishes on shell. In particular, \cite{Maxfield:2016mwh} showed that the on-shell partition function for saddles of the form \eqref{MM_wormhole} is
\begin{equation}
    I = - \frac{c}{3} \, (g-1)  \log(R_0) \ ,
\end{equation}
where $R_0$ is a parameter that is used to set the relative normalization of different saddles. Finally, the last piece was also mentioned in the original Maldacena-Maoz work \cite{Maldacena:2004rf}, and essentially is given by the contribution of two Einstein-Rosen bridges, suitably weighted by $1/Z^2(\beta)$. The reader should also notice that the dependence on $t-s$ could have been predicted a priori, since it is a symmetry of this saddle.  A similar result was also found in \cite{Stanford:2020wkf} for the case of JT gravity. Corrections to this result are the sum over images, which we have already argued does not affect much the result, and subleading contributions coming from different topologies.

It is instructive now to compare~\eqref{2D_gravity_expectations} with the corresponding results found in Section~\ref{sec:Correl_funct_quantum_chaotic_systems}. In particular, the result~\eqref{2D_gravity_expectations} gives a prediction on the functions $F_{2}(E_{\beta}, \omega)$ and $H_{4}(E_{\beta}, \omega)$. They can be both found employing a Mellin-Barnes transform
\begin{equation}
    \left( \frac{\pi}{\beta} \right)^{2 \Delta} \, \cosh^{-2 \Delta} \left( \frac{\pi t }{\beta} \right) = \int_{- \infty}^{\infty} \left( \frac{2 \pi}{\beta} \right)^{2 \Delta - 1} \frac{1}{\Gamma(2 \Delta)} \, \bigg| \Gamma \left(\Delta + \frac{i \omega \beta}{2 \pi} \right) \! \bigg|^2  \, \frac{e^{i \omega t}}{2 \pi} \, \de \omega 
\end{equation}
applied to the result~\eqref{avg_two_pt_gravity} for the mean and to the last contribution in~\eqref{2D_gravity_expectations}. This gives
\begin{equation}
    F_2(E_{\beta}, \omega)  =   \frac{2^{2 \Delta - 2} \pi^{2 \Delta - 2}}{\Gamma(2 \Delta) \beta^{2 \Delta - 1}}  \, \bigg| \Gamma \left(\Delta + \frac{i \omega \beta}{2 \pi} \right) \! \bigg|^2 = \frac{\beta^{2 \Delta - 1}}{2 \pi^{2 \Delta}} \, H_{4}(E_{\beta}, \omega) \ .
    \label{F2_H4_gravity}
\end{equation}
Let us also point out that the last two contributions in~\eqref{2D_gravity_expectations} have the same functional form at large $t,s$, up to $O(1)$ numbers which go to unity in the limit of light probes. This coincides with the expectations of Section~\ref{sec:Correl_funct_quantum_chaotic_systems}, where we have shown that they are both a Fourier transform of $H_{4}(E_{\beta}, \omega)$.

It is also interesting to compare the result~\eqref{2D_gravity_expectations} with the one obtained for Brownian motion in Section~\ref{BM_and_thermal_two_pt}. We find a similar functional structure, where contributions exponentially decaying as functions of $t + s$, and $|t-s|$. This qualitative match becomes quantitative in a double scaling limit where $\Delta \to 0$ and $\beta \to 0$ with 
\begin{equation}
    \gamma \equiv \frac{2 \pi \Delta}{\beta}  \qquad   \text{fixed} \ .
\end{equation}
This is the limit where we consider a very hot black hole and a particularly light scalar probe. Indeed, it is not hard to show that
\begin{equation}
    \lim_{\Delta \to 0}\cosh \left( \frac{\gamma \, t}{2 \Delta} \right)^{-2 \Delta} = e^{-\gamma \, |t|}  \ ,
\end{equation}
so that we can precisely match the autocorrelation function~\eqref{Brownian_two_pt} identifying the various parameters as
\begin{equation}
    v_0 \equiv \frac{\pi^{2 \Delta}}{\beta^{2 \Delta}} \ , \qquad  \gamma \equiv \frac{2\pi \Delta}{\beta} \ , \qquad \frac{g}{2 \gamma} \equiv \frac{\pi^{4 \Delta}}{Z^2(\beta) \, \beta^{4 \Delta}} \ .
\end{equation}
It is rather remarkable that the class of observables~\eqref{Two_pt_probe} computed in two-boundaries wormhole geometries suggests that the associated process belongs to the same universality class as Brownian motion (up to Gaussian order). The gravitational stochastic process however receives important non-Gaussian corrections considering wormholes with three or more boundaries \cite{Belin:2023efa, deBoer:2024kat}.

The upshot of this Section represents the main objective of this work, which is to argue that correlation functions in semiclassical gravity can be suitably interpreted as moments of a stochastic process. For such observables, gravity is then a stochastic description of a microscopic one. Interestingly, gravity computes moments of the stochastic process, rather than the stochastic process itself as in the case of Brownian motion and Section~\ref{sec:Correl_funct_quantum_chaotic_systems}. We will expand more on this point in Section~\ref{sec:Discussion}.

\subsection{Ensemble average vs expectation value}

In the previous Section we have been careful about using the expression {\it expectation value} rather than {\it ensemble average}. The difference between the two is subtle. The former refers to the expectation value over the probability distributions of the stochastic terms in the expression, which is a formal mathematical procedure. The latter refers to an operative protocol to take this expectation value, namely to realize the system a number of times. In this short paragraph, we point out that the two are rather distinct concept, so that an ensemble average is {\it not} needed in order to find expectation values. 

Clearly, fixing the operator $\cO$, the observable~\eqref{Two_pt_probe} is unique for every realisation of the system. At late times, it will show noisy thermal fluctuations, which can be smoothed out averaging over different realisation of the system. However, this is not the only way. A different one would consist in considering the class of observables
\begin{equation}
    G_{E}(t) =  \bra{E} e^{- \beta H/2} \, \cO(t) \, e^{- \beta H/2} \, \cO(0)  \ket{E}  \ ,
    \label{microcanonical_two_pt}
\end{equation}
where the eigenstate $\ket{E}$ is taken to belong to the microcanonical energy window of $E_{\beta}$. The observables~\eqref{microcanonical_two_pt} will all behave similarly to~\eqref{Two_pt_probe} in the fluctuations, and considering different ones we can perform an expectation value without resorting to an ensemble average. This idea is clearly realised also in the Brownian perspective, where the analogy stands preparing the system many time at the thermal energy $E_{\beta}$, but without being able to prepare it every time in the same microstate. 

From this point of view, an ensemble average seems a superfluous concept, and we believe {\it expectation value} is much better suited for this framework.

\subsection{Matryoshka correlators}
\label{sec:Gravity_matryoshkas}

Clearly, the autocorrelation function of $G_{\beta}(t)$ computed in~\eqref{2D_gravity_expectations} does not retain complete informations on the thermal fluctuations of the system. In particular, it captures the leading contribution of the fluctuations at late times as long as $|t-s|$ is small. However, in the limit of $|t-s|$ large, the result~\eqref{2D_gravity_expectations} becomes arbitrarily small. The physical reason is simple to explain: the correlation between the fluctuations at $t$ and at $s$ wash out in the limit $|t-s|$ large, and the autocorrelation decays. Again, we can easily see that this is inconsistent with a unitary microscopic description, but not with a stochastic interpretation. The signal $G_{\beta} (t) G_{\beta}(s)$, without taking the expectation value, at late times is a noisy signal of the order $1/Z^2(\beta)$. The fact that~\eqref{2D_gravity_expectations} vanishes just reflects the fact that, when $|t-s|$ is large itself, the value of the signal in $t$ is not correlated to $s$ anymore. To compute the size of the fluctuations of $G_{\beta} (t) G_{\beta}(s)$ in the limit $|t-s|$ large, we could consider the quantity $\mathbb E[G_\beta^2(t)] \, \mathbb E[G_\beta^2(s)]$, which is non-vanishing and captures the size of the fluctuations in that limit. However, a more refined quantity is the fourth moment $\mathbb E [ G_{\beta}(t)  G_{\beta}(s) G_{\beta}(p)  G_{\beta}(q) ]$. Clearly, when $t = p$ and $s = q$, in the limit $|t-s|$ large, this moment reduces to $\mathbb E [ G_{\beta}^2(t)  G_{\beta}^2(s) ] \approx \mathbb E[G_\beta^2(t)] \, \mathbb E[G_\beta^2(s)] + \dots$, which is non-vanishing. 

This structure amounts to consider non-Gaussian corrections, which are not visible at the level of the two-boundary wormhole. For the fourth moment aforementioned, these corrections in holography correspond to three- and four-boundary wormholes. From a stochastic perspective, such geometries describe correlations between different instances of the signal, that are not captured by a Gaussian process. For this reason, they contain more finely grained information. This structure is particularly intriguing in the idea of bootstrapping microscopic information of quantum gravity from higher and higher moments of a stochastic process. We will come back to it in Section~\ref{sec:Discussion}. Let us however remark that a universal mechanism seems to be at play, since a similar mechanism was noted to happen for generic systems respecting ETH \cite{Foini-Pappalardi-Kurchan_free_prob}, while other works in the literature proposed matrix/tensor models from non-Gaussian corrections \cite{Jafferis:2022uhu, Jafferis:2022wez, Belin:2023efa} that can possibly be used as a non-perturbative definition of a gravitational path integral. 

The physical insight that this recursive structure suggests is that in a mesoscopic description there is no fixed moment that is able to capture universally all microscopic fluctuations. On the other hand, one is forced to consider the whole series of moments to be consistent with a unitary microscopic theory. 

On a conceptual level, this can be seen as a necessary outcome of promoting a deterministic description to a probabilistic one, the price to pay in a sense. Once we understand that a finer amount of information is hidden in a higher moment, then we are forced to consider the whole series of higher moments, as we attempted to convey with the Matryoshka analogy. Certainly, discovering a secret doll inside a bigger one, is the best clue to look for another one, and another one, and another one...

\section{Mesoscopic CFTs} \label{sec:Mesoscopic_CFTs}

We now turn our attention to a microscopic derivation of the gravitational results found in Section~\ref{sec:different_moments}. The general idea relies on the OPE expansion of thermal correlation functions of `simple' CFT primaries. This expansion will depend on OPE data which, in a similar fashion to Section~\ref{BM_and_thermal_two_pt}, will be interpreted as probability distributions, which imparts a stochastic nature to the correlation function itself.

\subsection{Lightning review of holographic CFTs}

The microscopic system under consideration is a two-dimensional CFT which has a large central charge $c$, and a gap in the operator spectrum. In particular, it has only a sparse amount of primary operators with scaling dimensions less that $(c-1)/24$, which will constitute our probes in a thermal background \cite{Hartman:2014oaa}. We will refer to these operators as {\it light}, {\it i.e.} the `simple' operators mentioned above, while all the others will be {\it heavy} operators.

Using these assumptions, it is possible to prove that all CFTs belonging to this class display a universal set of features. In particular, the spectrum of heavy primary operators follows a Cardy distribution \cite{Hartman:2014oaa}, 
\begin{equation}
    \rho(h, \bar h) \approx e^{2 \pi \sqrt{\frac{c}{6} \left( h - \frac{c}{24} \right)} + 2 \pi \sqrt{\frac{c}{6} \left( \bar h - \frac{c}{24} \right)}} = \rho_0(h) \, \rho_0(\bar h) \ ,
    \label{Cardy_density}
\end{equation}
while OPE coefficients follow the asymptotic behavior \cite{Collier:2019weq}
\begin{equation}
    |C_{pqr}|^2 \sim C_0(p, q, r) \, C_0(\bar p, \bar q, \bar r) \ ,
    \label{distribution_of_OPE_coeff}
\end{equation}
where $C_0$ is related to the DOZZ formula \cite{DORN1994375, ZAMOLODCHIKOV1996577}. It is important to stress that~\eqref{distribution_of_OPE_coeff} is only valid when at least one operator is heavy. On the other hand, we can consider light operators to be free fields, so that all light-light-light three point functions vanish except
\begin{equation}
    C_{pp\mathbbm 1} \equiv 1 \ ,
\end{equation}
which is the same for all operators in the spectrum, setting the normalisation of two point functions. Beside these kinematic constraints, the dynamics is set by the generators of time translations, which for a compact CFT of length $2\pi$ is
\begin{equation}
    H = L_0 + \bar L_0 - \frac{c}{12} \ .
    \label{H_and_S_compact_CFT}
\end{equation}
This implies that primary operators and descendants are eigenstates of the Hamiltonian~\eqref{H_and_S_compact_CFT}. From the Cardy density~\eqref{Cardy_density} and the Hamiltonian~\eqref{H_and_S_compact_CFT}, it is also possible to study the thermodynamics of the system. The thermodynamic energy and entropy can be found considering the saddlepoints of the thermal partition function. Using the Cardy density~\eqref{Cardy_density}, we obtain
\begin{equation}
    E_{\beta} = \frac{\pi^2 c}{3 \beta^2} \qquad \text{and} \qquad S_{\beta} = \frac{2 \pi^2 c}{3 \beta} \ .
\end{equation}
We finish this paragraph commenting on the spectral width. From the Cardy density, it is not difficult to compute $\delta E^2 = \langle E^2 \rangle_\beta - E_{\beta}^2 = 8 \pi^2 c / 3 \beta^3 + \dots$ in a saddlepoint approximation. As shown, we find that it scales linearly with $c$, making the spectral width parametrically large in the large-$c$ limit.

All in all, in this short review of holographic CFTs we have shown that such systems meet all the necessary requirements laid out at the beginning of Section~\ref{BM_and_thermal_two_pt} to be interpreted mesoscopically.

\subsection{CFTs and mesoscopic descriptions}

In analogy with~\eqref{Two_pt_probe}, we consider thermal correlation functions for CFT primary operators. It is particularly convenient to write the correlation function in terms of CFT variables. Introducing the complex coordinate $z = \theta + i t_E$ on the torus, where $\theta \simeq \theta + 2 \pi$ and $t_E \simeq t_E + \beta$, by conformal invariance physical observables will depend only on the quantity $\tau  = i \beta / 2\pi$. The thermal correlation function~\eqref{Two_pt_probe} can then be written as
\begin{equation}
    G_{\beta}(z, \bar z) =  \frac{1}{Z(\tau, \bar \tau)} \, \Tr \left[ e^{2 \pi i \tau (L_0 - c/24) - 2 \pi i \bar \tau (\bar L_0 - c/24)} \, \cO(z, \bar z) \, \cO(0) \right] 
    \label{torus_correlator_trace}
\end{equation}
where the sum runs over all operators, primaries and descendants, and the normalisation $Z(\tau, \bar \tau)$ is the thermal partition function
\begin{equation}
    Z(\tau, \bar \tau) = \Tr \left[ e^{2 \pi i \tau (L_0 - c/24) - 2 \pi i \bar \tau (\bar L_0 - c/24)} \right] = \Tr \left[ e^{- \beta H} \right] \ .
\end{equation}
Taking\footnote{Because of the real-time evolution, it is not surprising that $\bar z$ is {\it not} the complex conjugate of $z$.} $z = (i \beta /2) - t$ and $\bar z = (- i \beta /2) + t$, it is a simple exercise to show that~\eqref{torus_correlator_trace} reduces to the thermal two-point function~\eqref{Two_pt_probe}. To exploit conformal invariance, it is then customary to rewrite the sum into torus conformal blocks, so that the sum only depends on the OPE data. In the {\it necklace channel} \cite{Cho:2017oxl, Collier:2019weq}, this is done as
\begin{equation}
    G_{\beta} (z, \bar z) = \frac{1}{Z(\tau, \bar \tau)} \sum_{p_1, p_2} |C_{\cO p_1 p_2}|^2  \, \cF_{\rm N} (p_1,p_2| \tau, z) \bar \cF_{\rm N} (\bar p_1, \bar p_2|\bar \tau, \bar z) \ ,
    \label{two_pt_fn_s_channel}
\end{equation}
where the sum is over Virasoro primaries which are being exchanged in the OPE. These are all heavy, since the light operators do not interact. A graphical representation can be found in Figure~\ref{OPE_vs_Neclace_channel}.

\begin{figure}[t]
\centering
    \begin{tikzpicture}[scale = 0.8]

        \begin{scope}[scale=2, every node/.append style={transform shape}]
            \draw[] (-1.9,-0.15) node {$\displaystyle \sum_{p_1, p_2}$};
        \end{scope}

        \draw[very thick] (0,0) circle (1);
        \draw[very thick] (-2,0) -- (-1,0);
        \draw[very thick] (1,0) -- (2,0);

        \draw[] (-2,0) node[anchor = east] {$\cO$};
        \draw[] (2,0) node[anchor = west] {$\cO$};
        \draw[] (0,1) node[anchor = south] {$p_1$};
        \draw[] (0,-1) node[anchor = north] {$p_2$};

        \draw[] (0,-2.5) node {Necklace};
        \draw[] (0,-3) node {channel};

        \begin{scope}[scale=2, every node/.append style={transform shape}]
            \draw[] (2.45,-0.15) node {$\displaystyle \sum_{p_1', p_2'}$};
        \end{scope}

        \begin{scope}[scale=2, every node/.append style={transform shape}]
            \draw[] (1.7,0) node {$ = $};
        \end{scope}

        \begin{scope}[scale=2, every node/.append style={transform shape}]
            \draw[] (-3.5,0.02) node {$G_{\beta}(z, \bar z) \; = $};
        \end{scope}

        \begin{scope}[xshift=10]

            \draw[very thick] (7,0) circle (1);
            \draw[very thick] (7,1.7) -- (7,1);
            \draw[very thick] (6,2) -- (7,1.7);
            \draw[very thick] (8,2) -- (7,1.7);
    
            \draw[] (5.6,2.1) node {$\cO$};
            \draw[] (8.4,2.1) node {$\cO$};
            \draw[] (7,-1.35) node {$p_2'$};
            \draw[] (7.45,1.4) node {$p_1'$};
    
            \draw[] (7,-2.5) node {OPE};
            \draw[] (7,-3) node {channel};
        \end{scope}

        \draw[white] (0,-3.5);
        
    \end{tikzpicture}
    \caption{Different channels for the thermal two point function.}
    \label{OPE_vs_Neclace_channel}
\end{figure}

As in Section~\ref{BM_and_thermal_two_pt}, the mesoscopic description arises considering the OPE data as probability distributions, such that \cite{Belin:2020hea, Chandra:2022bqq} 
\begin{equation}
    \mathbb E[C_{pqr}] = 0 \ , \qquad \qquad \mathbb E \big[|C_{pqr}|^2 \big] = C_0(p, q, r) \, C_0(\bar p, \bar q, \bar r) \ .
    \label{mean_variance_three_pt_functions}
\end{equation}
One can see this class of CFTs as an ensemble, and average over the various draws of the distribution. Quite remarkably, considering uncorrelated Gaussians, it is possible to find agreement with different classes of two-boundary wormhole contribution in gravity \cite{Belin:2020hea, Chandra:2022bqq}, while considering necessary non-Gaussian corrections leads to a tensor model which resembles a discrete version of three-dimensional gravity \cite{Belin:2023efa}. Some Readers may object that, in most of the cases, a specific realisation of this probability distribution will not be a consistent CFT, not satisfying all crossing and modular constraints, which led the authors of \cite{Belin:2023efa} to define the notion of an approximate CFT. We would like to point out in the present context that this is a standard practice in statistical mechanics. Taking as an example a classical interacting theory, the statistical mechanics approach integrates the phase space {\it e.g.} in the whole microcanonical window, rather than restricting only to points in the phase space that can be connected by collisions where energy and momentum are conserved. The underlying claim at play here is that the emergent universal phenomena are not sensitive enough to detect microscopic inconsistencies.

We are now in the position to consider moments of such stochastic processes. For this purpose, it is useful to consider a different OPE expansion, namely the {\it OPE channel}, in which 
\begin{multline}
    G_{\beta} (z, \bar z) = \frac{1}{Z(\tau, \bar \tau)} \, \bigg[ \sum_{p'}  \cF_{\rm OPE} (\mathbbm 1, p'| \tau, v) \, \bar \cF_{\rm OPE} ( \mathbbm 1, \bar p'| \bar \tau, \bar v) \, \, +   \\
    \sum_{p_1', p_2' \neq \mathbbm 1} C_{\cO \cO p_1'} C_{p_1' p_2' p_2'} \, \cF_{\rm OPE} (p_1', p_2'| \tau, v) \, \bar \cF_{\rm OPE} ( \bar p_1', \bar p_2'| \bar \tau, \bar v)  \bigg] \ ,
    \label{Mesoscopic_expansion}
\end{multline}
where the OPE is obtained first fusing the light primaries, and then evolving around the thermal circle. The modulus $v$ is related to $z$, and precise relation can be found {\it e.g.} in \cite{Yellow_Book, Cho:2017oxl}. This channel is particularly convenient since the second term in the RHS of~\eqref{Mesoscopic_expansion} has mean zero, and thus
\begin{multline}
    \mathbb E[G_{\beta} (z, \bar z)] = \frac{1}{Z(\tau, \bar \tau)} \sum_{p_1, p_2} \mathbb E \Big[|C_{\cO p_1 p_2}|^2 \Big]  \, \cF_{\rm N} (p_1,p_2| \tau, z) \bar \cF_{\rm N} (\bar p_1, \bar p_2| \bar \tau, \bar z) \\
    = \frac{1}{Z(\tau, \bar \tau)} \,  \sum_{p'}  \cF_{\rm OPE} (\mathbbm 1, p'| \tau, v) \, \bar \cF_{\rm OPE} ( \mathbbm 1, \bar p'| \bar \tau, \bar v) = \frac{\pi^{2 \Delta}}{\beta^{2 \Delta}} \cosh^{- 2 \Delta} \left( \frac{\pi t}{\beta} \right) + \dots \ ,
\end{multline}
where in the last passage we have employed results on the identity block exchange \cite{Fitzpatrick:2015zha, Anous:2019yku, Gerbershagen:2021yma}. Unsurprisingly, this matches the gravity expectation for the mean~\eqref{avg_two_pt_gravity}, where in both expressions we have neglected the sum over images, or equivalently over geodesics that wrap around the black hole. 

As before, we can also go further and consider {\it e.g.} the second moment of the correlation function, namely $\mathbb E[G_{\beta} (z, \bar z) G_{\beta} (w, \bar w)]$. The complex parameter $w$ is the modulus corresponding to the Lorentzian time $s$, which was also used in~\eqref{2D_gravity_expectations}.
In this case, a connection with the results of Section~\ref{BM_and_thermal_two_pt} is more straightforward in the necklace channel. In Appendix~\ref{sec:autocorrelation_in_CFT} we show that the result is
\begin{multline}
    \mathbb E[G_{\beta} (z, \bar z) G_{\beta} (w, \bar w)] \, \approx \,  \, \mathbb E[G_{\beta} (z, \bar z)] \, \mathbb E[G_{\beta} (w, \bar w)] \, + \frac{1}{Z^2(\tau, \bar \tau)} \, G_{(\tau, - \tau)}^L (z, - w) \, G_{(- \bar \tau, \bar \tau)}^L (- \bar z, \bar w) \, \\
    + \frac{1}{Z^2(\tau, \bar \tau)} \, G_{(\tau, -\tau)}^L (z, w) \, G_{(-\bar \tau, \bar \tau)}^L (\bar w, \bar z)\ ,
    \label{two_pt_fn_thermal}
\end{multline}
where $G_{(\tau, \bar \tau)}^L (z, \bar z)$ is the thermal Liouville two point function. This result is consistent with the findings of \cite{Chandra:2022bqq}, which showed that any two-boundary wormhole contribution in gravity will have the form~\eqref{two_pt_fn_thermal}, and is also in a one-to-one correspondence with the gravity expectation~\eqref{2D_gravity_expectations} found in Section~\ref{sec:different_moments}.

\subsection*{A note on ramp and plateau}

It is worth noting that in the above discussion we have completely neglected connected correlations of the spectral density $\overline{\rho(E) \rho(E')} - \overline{\rho(E)} \, \,  \overline{\rho(E')}$, which is known are connected with a ramp-plateau behavior of correlation functions \cite{Altland:2021rqn, Bouverot-Dupuis:2024nxk} in chaotic systems. For instance, a computation of the ramp-plateau structure in JT gravity was performed in \cite{Saad:2019pqd}, where it was shown that {\it e.g.} the ramp arises from geometries with handles. However, in the present context we are interested on fluctuations over the average, which is more in line with \cite{Stanford:2020wkf}. Moreover, the ramp-plateau behavior appears later than $t \sim S$, which is the timescale at which the leading wormhole contribution start being dominant.

\subsection{Vacuum correlation functions}

Similar considerations also suggest to treat generic $n$-point functions in holographic CFTs, such as correlation functions on the ground state, as stochastic processes. Let us show it in the simplest possible example, namely the four-point function 
\begin{equation}
    G_{(4)}(x, \bar x) = \langle \cO_i(0) \cO_j(1) \cO_j(x, \bar x) \cO_i(\infty)  \rangle \ .
\end{equation}
This can be expanded into two channels that are commonly called the $s$-channel and the $t$-channel. The former is
\begin{equation}
    G_{(4)}(x, \bar x) = \sum_{p} | C_{ijp} |^2 \, \cF_{\rm s} (p|x) \bar \cF_{\rm s} (\bar p| \bar x) \ ,
    \label{Vacuum_4pt_s-channel}
\end{equation}
while the $t$-channel is
\begin{equation}
    G_{(4)}(x, \bar x) = \cF_{\rm t} (\mathbbm 1|1- x) \bar \cF_{\rm t} (\mathbbm 1| 1 - \bar x) + \sum_{p \neq \mathbbm 1} C_{iip} C_{pjj} \, \cF_{\rm t} (p|1- x) \bar \cF_{\rm t} (\bar p| 1 - \bar x) \ .
    \label{Mesoscopic_4pt_function}
\end{equation}
In~\eqref{Mesoscopic_4pt_function} we have divided the contribution coming from the identity block exchange from the contribution coming from heavy operators exchanges. The two expansions~\eqref{Vacuum_4pt_s-channel} and~\eqref{Mesoscopic_4pt_function} can be related by a crossing transformation, but we find the $t$-channel expression more convenient for our analysis. In particular, the identity-block contribution is the mean of the process,
\begin{equation}
    \mathbb E[G_{(4)}(x, \bar x)] = \cF_{\rm t} (\mathbbm 1|1- x) \bar \cF_{\rm t} (\mathbbm 1| 1 - \bar x)  \ ,
    \label{Mesoscopic_4pt_function_mean}
\end{equation}
while the second contribution on the RHS is a mean-zero stochastic process. We can also compute the second moment, finding
\begin{multline}
    \mathbb E[G_{(4)}(x_1, \bar x_1) G_{(4)}(x_2, \bar x_2)] = \mathbb E[G_{(4)}(x_1, \bar x_1)] \, \mathbb E[G_{(4)}(x_2, \bar x_2)] \, + \\
    \left| \int_{\frac{c-1}{24}}^{\infty} \de h_p \, \rho(h_p) \, C_0(h_1, h_1, h_p) \, C_0(h_2, h_2, h_p) \, \cF^{11}_{22} (p|1- x_1) \, \cF^{11}_{22} (p|1- x_2) \right|^2 \\
    = \mathbb E[G_{(4)}(x_1, \bar x_1)] \, \mathbb E[G_{(4)}(x_2, \bar x_2)] + G_{(4)}^L(x_1, x_2) G_{(4)}^L(\bar x_2, \bar x_1) \ ,
\end{multline}
where $G_{(4)}^L(x_1, x_2)$ is a Liouville vacuum four-point function, consistent with \cite{Chandra:2022bqq}.

\section{Discussion and future directions} \label{sec:Discussion}

The main objective of this paper has been to establish an interpretation of semiclassical gravity in Anti-de Sitter as a mesoscopic description of a microscopic quantum theory of gravity. We have explained how wormholes between tori with matter insertions can be thought of moments of such a stochastic description of the microscopic correlation function, and shown a quantitative agreement with Brownian motion, up to the second moment. 

The similarities with the latter are quite remarkable. A thermal correlation function in gravity cannot be taken literally as a microscopic calculation because it lacks the necessary thermal fluctuations at late times, in the same way as a particle that forever slows down because of a dynamical viscosity in a fluid is incompatible with basic requirements of thermodynamics. The calculation of the second moment of the effective quantity, whether it be a correlation function in gravity or the square of the velocity for a particle, recovers such fluctuations in a probabilistic way. It is striking how such seemingly different systems share the same underlying physical phenomenology. In this sense however, semiclassical gravity is unique in the fact that this probabilistic description is not forced `by hand', as it happens in promoting a deterministic dissipative system to Brownian motion, but it appears naturally as the simplest way to explain wormhole geometries in the gravitational path integral.

On a more technical level, the outcome of this study sharpens some ideas that have been discussed in the literature regarding the rôle of wormholes in semiclassical theories of gravity \cite{Saad:2018bqo, Altland:2020ccq, Almheiri:2019hni, Belin:2020hea, Chandra:2022bqq, Belin:2023efa}. In particular, our study employed the so-called {\it Euclidean wormholes} to study the loss-and-recovery of thermal fluctuations in gravity, similarly to \cite{Stanford:2020wkf}, but interpreting the outcome in the fashion of stochastic processes. Interestingly, in the way we have framed the discussion, any notion of ensemble averaging seems superfluous, and justifies our use of the word {\it expectation value} in a realization of the system, rather than considering an ensemble of them. 

An interesting perspective that we have already mentioned is that, in this framework, semiclassical gravity can be considered as a moment generating function for the underlying stochastic process. This perspective, suggested by the results of Section~\ref{sec:Gravity_expectations}, is complementary to the one described in Sections~\ref{BM_and_thermal_two_pt} and~\ref{sec:Mesoscopic_CFTs}, where a mesoscopic description arises from the microscopic one specifying a probability distribution. This can also be seen as a new interpretation of AdS/CFT, where a semiclassical theory of gravity is dual to a mesoscopic description of a holographic CFT. From this viewpoint, it's possible to establish a mesoscopic holographic duality where semiclassical gravity, {\it i.e.} a moment generating function, is dual to a mesoscopic CFT, {\it i.e.} a probability distribution. 

Even though the two sides of the duality should be equivalent, the precise relation between specifying the moments and specifying a probability distribution is still an open problem in mathematics. An interesting direction to explore could exploit the bootstrap methods developed for matrix models \cite{Lin:2020mme} to moments of stochastic processes, demanding a positive definite measure\footnote{For a recent work computing higher moments of the probability distribution, see \cite{deBoer:2024kat}.}. Exploring whether such constraints are strong enough to give access to fine microscopic data is an additional intriguing direction to pursue. On one hand, considering higher and higher moments, it seems that we are probing finer and finer observables. On the other, such moments still depend on a handful of macroscopic parameters, and it thus seems that a wealth of microscopic theory could be described by the same universal mesoscopic description. The problem is far from being understood.

Another interesting direction to explore is the connection between this work and a recent line of research that considers random tensor models\footnote{For random matrix models in the context of JT gravity, see \cite{Jafferis:2022uhu, Jafferis:2022wez}.} \cite{Belin:2023efa}
\begin{equation}
    \mathcal Z = \int D[L_0, \bar L_0] \, D[C] \; e^{- a V(L_0, \bar L_0, C)} \ ,
    \label{Tensor_model}
\end{equation}
as a non-perturbative definition of gravity. The potential $V(\cO, C)$ is found imposing CFT crossing conditions {\it on average}, and the interplay between propagators and interaction vertices generate geometrical structures. From our perspective, the integrand in~\eqref{Tensor_model} specifies the probability distribution that determines the stochastic process at play, with non-Gaussian corrections. It would be interesting to see how a geometric notions for the observables~\eqref{generic_thermal_corr_funct} arises in this context, and if it can be connected with more familiar Fokker-Planck path integrals. An interesting clue could be given by the OPE channel \eqref{Mesoscopic_expansion}, which naturally splits the mean of the process and the fluctuating part. The former is clearly a self-averaging quantity, that has a clear geometrical description in terms of a BTZ black hole with a propagating particle. The latter is not self-averaging, and its bulk dual is rather unclear. However, one of the outcomes of this work is that such a fluctuating part can be paired with other ones in order to give a clear geometric quantity, the wormholes. For this reason, we depict it as in Figure \ref{fig:Non-self_averaging}. This discussion also seems related to the concept of {\it half wormholes}, introduced in \cite{Saad:2021rcu}. It would be interesting to see if this connection can be made more precise, especially improving the toy model studied in \cite{Saad:2021rcu}, perhaps borrowing tools from \cite{DiUbaldo:2023qli}. 

\begin{figure}[t]
\centering
    \begin{tikzpicture}[scale=0.9]

        \draw[] (-2.5, -1.25) node {\Large $+$};

        \draw[] (-7.5, -1.25) node {\Large $=$};

        \draw[] (-9.2, -1.2) node {\Large $G_{\beta}(t)$};

        
        \begin{scope}[yshift = -1.25cm]  

        \draw[red,fill=red] (-4.2, -0.3) circle (2pt) ;
        \draw[red,fill=red] (-5.8, -0.3) circle (2pt) ;

        \draw[color = red, thick] plot [smooth, tension=1] coordinates { (-4.2, -0.3)  (-5, -0.55)  (-5.8, -0.3)};

        \draw[very thick] (-5,0) ellipse (1.5 and 1);
        \draw[very thick] plot [smooth, tension=1] coordinates { (-5.5, 0.1)  (-5, -0.1)  (-4.5, 0.1)};
        \draw[very thick] plot [smooth, tension=1] coordinates { (-5.3, -0.04)  (-5, 0.11)  (-4.7, -0.04)};
        \end{scope}


        \begin{scope}[yshift = 1.25cm]  
        \draw[very thick] (0,-2.5) ellipse (1.5 and 1);
        \draw[very thick] plot [smooth, tension=1] coordinates { (-0.5, -2.4)  (0, -2.6)  (0.5, -2.4)};
        \draw[very thick] plot [smooth, tension=1] coordinates { (-0.3, -2.54)  (0, -2.39)  (0.3, -2.54)};

        \draw[red,fill=red] (0.8, -2.8) circle (2pt) ;
        \draw[red,fill=red] (-0.8, -2.8) circle (2pt) ;

        \draw[color = red, thick] plot [smooth, tension=1] coordinates { (0.75, -1.1)  (0.73, -1.85)  (0.8, -2.8)};
        
        \draw[color = red, thick] plot [smooth, tension=1] coordinates { (-0.8, -0.95)  (-0.73, -1.85)  (-0.8, -2.8)};

        \fill[color = Periwinkle, opacity = 0.2] plot [smooth, tension=1] coordinates { (-1.5, -1)  (-1.4, -1.75)  (-1.5, -2.5)} -- (-1.5,-2.5) arc (180:360:1.5 and 1) -- plot [smooth, tension=1] coordinates { (1.5, -2.5) (1.4, -1.75) (1.5, -1) } -- plot [smooth, domain=1.5:-1.5, smooth, variable=\x] ({\x}, {-0.13*sin(12*90*\x) - 1}) -- cycle;

        \draw[color = Periwinkle, thick] plot [smooth, tension=1] coordinates { (-1.5, -2.5) (-1.4, -1.75) (-1.5, -1)};
        \draw[color = Periwinkle, thick] plot [smooth, tension=1] coordinates { (1.5, -2.5)  (1.4, -1.75)  (1.5, -1)};

        \draw[color = Periwinkle, thick] plot [smooth, domain=1.5:-1.5, smooth, variable=\x] ({\x}, {-0.13*sin(12*90*\x) - 1}) ;
        \end{scope}

    \draw[] (0, -3);
        
    \end{tikzpicture}
    \caption{A bulk representation of the correlation function \eqref{generic_thermal_corr_funct} in the OPE channel \eqref{Mesoscopic_expansion}. The first contribution in the RHS is the mean and has a clear bulk dual. The second contribution is a fluctuating quantity that can be paired with others to form wormholes. }
    \label{fig:Non-self_averaging}
\end{figure}

Somewhat remarkably, the formulation laid down in Section~\ref{sec:Mesoscopic_CFTs} seems to be closely connected to the Kosambi, Karhunen and Loève (KKL) theorem \cite{Karhunen47, Loeve48}. The KKL theorem states that for any stochastic process $X(t)$, it exists a basis of functions $f_n(t)$ such that the coefficients $c_n$ of the expansion
\begin{equation}
    X(t) = \sum_n c_n \, f_n (t) 
\end{equation}
are uncorrelated random variables, thus respecting $\mathbb E[c_n c_m] = \lambda_n \, \delta_{mn}$. In this work, we have essentially shown that for the stochastic processes described by gravity in three dimensions, such a basis is the set of conformal blocks, and that the variance $\lambda_n$ is related to the DOZZ formula as in \eqref{mean_variance_three_pt_functions}. It would be particularly interesting to delve deeper into the theory of stochastic processes, to see if there could be interesting lessons to be learnt about gravity.

It is interesting to note that the issue of Brownian motion in (applied) AdS/CFT has been studied in \cite{deBoer:2008gu,Son:2009vu}, but in a rather different context. There a fundamental charge was inserted into the boundary theory at finite temperature, corresponding to a fundamental string stretching from the charge down to the black-hole horizon. The Hawking radiation emerging from the black hole then manifests itself as a stochastic noise term in a \emph{boundary} Langevin equation for the fundamental charge. By contrast, in the present paper, we point out that bulk (semi-classical) gravity itself can be seen as a stochastic process, akin to Brownian motion, owing to the presence of multi-boundary geometries contributing to the path integral. It would be interesting to see if the world-sheet interpretation of \cite{deBoer:2008gu,Son:2009vu}, may in fact also be interpreted as coming from "worldsheet wormholes", as in \cite{Sonner:2013mba,Jensen:2013ora}, or the more recent \cite{Post:2022dfi,Altland:2022xqx}, which give a world-sheet interpretation of the AdS$_2$ bulk wormhole story.

Finally, it would also be interesting to re-visit the results presented in this work from the point of view of open quantum systems, along the lines of \cite{Pelliconi:2023ojb, Hernandez-Cuenca:2024pey}. In particular, our study suggests that taking expectation values of noisy observables can be reinterpreted as a non-unitary evolution that decoheres the system. It is quite interesting that such non-unitary evolution does not arise because of the influence of an external field, but seems to be intrinsic in the low energy description. This is reminiscent of the lore that {\it quantum chaotic systems are their own bath}. A more systematic study of this mechanism in gravity could prove rewarding.

\vspace{1cm}

\noindent {\Large \bf Acknowledgments}

\vspace{0.3cm}
\noindent We would like to thank Tom Hartman, Kristan Jensen, Adrián Sánchez-Garrido, Douglas Stanford and Cynthia Yan for helpful conversations and comments. The research of P.P. and J.S. is supported in part by the Fonds National Suisse de la Recherche Scientifique (Schweizerischer Nationalfonds zur Förderung der wissenschaftlichen Forschung) through Project Grants 200021\_215300, NCCR51NF40-141869 The Mathematics of Physics (SwissMAP). The research of H.V. is supported by NSF grant PHY-2209997.

\bibliographystyle{utphys}
\bibliography{extendedrefs}

\providecommand{\href}[2]{#2}\begingroup\raggedright\begin{thebibliography}{10}

\bibitem{Zinn-Justin_book}
J.~Zinn-Justin,
  \href{http://dx.doi.org/10.1093/acprof:oso/9780198509233.001.0001}{{\em
  {Quantum Field Theory and Critical Phenomena}}}.
\newblock Oxford University Press, 2002.

\bibitem{PhysRevLett_52_1}
O.~Bohigas, M.~J. Giannoni, and C.~Schmit, ``Characterization of chaotic
  quantum spectra and universality of level fluctuation laws,''
  \href{http://dx.doi.org/10.1103/PhysRevLett.52.1}{{\em Phys. Rev. Lett.}
  {\bfseries 52} (1984) 1--4}.

\bibitem{BGS_laplacian}
O.~Bohigas, M.~Giannoni, and C.~Schmit, ``Spectral properties of the laplacian
  and random matrix theories,''
  \href{http://dx.doi.org/10.1051/jphyslet:0198400450210101500}{{\em J.
  Physique Lett.} {\bfseries 45} no.~21, (1984) 1015--1022}.

\bibitem{Altland_Zirnbauer_classification}
A.~Altland and M.~R. Zirnbauer, ``Nonstandard symmetry classes in mesoscopic
  normal-superconducting hybrid structures,''
  \href{http://dx.doi.org/10.1103/PhysRevB.55.1142}{{\em Phys. Rev. B}
  {\bfseries 55} (1997) 1142--1161}.

\bibitem{No_Hair_1}
W.~Israel, ``Event horizons in static vacuum space-times,''
  \href{http://dx.doi.org/10.1103/PhysRev.164.1776}{{\em Phys. Rev.} {\bfseries
  164} (1967) 1776--1779}.

\bibitem{No_Hair_2}
W.~Israel, ``Event horizons in static electrovac space-times,''
  \href{http://dx.doi.org/10.1007/BF01645859}{{\em Communications in
  Mathematical Physics} {\bfseries 8} no.~3, (1968) 245--260}.

\bibitem{No_Hair_3}
B.~Carter, ``Axisymmetric black hole has only two degrees of freedom,''
  \href{http://dx.doi.org/10.1103/PhysRevLett.26.331}{{\em Phys. Rev. Lett.}
  {\bfseries 26} (1971) 331--333}.

\bibitem{Bekenstein:1972tm}
J.~D. Bekenstein, ``{Black holes and the second law},''
  \href{http://dx.doi.org/10.1007/BF02757029}{{\em Lett. Nuovo Cim.} {\bfseries
  4} (1972) 737--740}.

\bibitem{Bekenstein:1973ur}
J.~D. Bekenstein, ``{Black holes and entropy},''
  \href{http://dx.doi.org/10.1103/PhysRevD.7.2333}{{\em Phys. Rev. D}
  {\bfseries 7} (1973) 2333--2346}.

\bibitem{Hawking:1974rv}
S.~W. Hawking, ``{Black hole explosions},''
  \href{http://dx.doi.org/10.1038/248030a0}{{\em Nature} {\bfseries 248} (1974)
  30--31}.

\bibitem{Hawking:1975vcx}
S.~W. Hawking, ``{Particle Creation by Black Holes},''
  \href{http://dx.doi.org/10.1007/BF02345020}{{\em Commun. Math. Phys.}
  {\bfseries 43} (1975) 199--220}. [Erratum: Commun.Math.Phys. 46, 206 (1976)].

\bibitem{Strominger:1996sh}
A.~Strominger and C.~Vafa, ``{Microscopic origin of the Bekenstein-Hawking
  entropy},'' \href{http://dx.doi.org/10.1016/0370-2693(96)00345-0}{{\em Phys.
  Lett. B} {\bfseries 379} (1996) 99--104},
  \href{http://arxiv.org/abs/hep-th/9601029}{{\ttfamily arXiv:hep-th/9601029}}.

\bibitem{Cotler:2016fpe}
J.~S. Cotler, G.~Gur-Ari, M.~Hanada, J.~Polchinski, P.~Saad, S.~H. Shenker,
  D.~Stanford, A.~Streicher, and M.~Tezuka, ``{Black Holes and Random
  Matrices},'' \href{http://dx.doi.org/10.1007/JHEP05(2017)118}{{\em JHEP}
  {\bfseries 05} (2017) 118}, \href{http://arxiv.org/abs/1611.04650}{{\ttfamily
  arXiv:1611.04650 [hep-th]}}. [Erratum: JHEP 09, 002 (2018)].

\bibitem{Saad:2018bqo}
P.~Saad, S.~H. Shenker, and D.~Stanford, ``{A semiclassical ramp in SYK and in
  gravity},'' \href{http://arxiv.org/abs/1806.06840}{{\ttfamily
  arXiv:1806.06840 [hep-th]}}.

\bibitem{Saad:2019lba}
P.~Saad, S.~H. Shenker, and D.~Stanford, ``{JT gravity as a matrix integral},''
  \href{http://arxiv.org/abs/1903.11115}{{\ttfamily arXiv:1903.11115
  [hep-th]}}.

\bibitem{Saad:2019pqd}
P.~Saad, ``{Late Time Correlation Functions, Baby Universes, and ETH in JT
  Gravity},'' \href{http://arxiv.org/abs/1910.10311}{{\ttfamily
  arXiv:1910.10311 [hep-th]}}.

\bibitem{Cotler:2020ugk}
J.~Cotler and K.~Jensen, ``{AdS$_{3}$ gravity and random CFT},''
  \href{http://dx.doi.org/10.1007/JHEP04(2021)033}{{\em JHEP} {\bfseries 04}
  (2021) 033}, \href{http://arxiv.org/abs/2006.08648}{{\ttfamily
  arXiv:2006.08648 [hep-th]}}.

\bibitem{Altland:2020ccq}
A.~Altland and J.~Sonner, ``{Late time physics of holographic quantum chaos},''
  \href{http://dx.doi.org/10.21468/SciPostPhys.11.2.034}{{\em SciPost Phys.}
  {\bfseries 11} (2021) 034}, \href{http://arxiv.org/abs/2008.02271}{{\ttfamily
  arXiv:2008.02271 [hep-th]}}.

\bibitem{Jafferis:2022uhu}
D.~L. Jafferis, D.~K. Kolchmeyer, B.~Mukhametzhanov, and J.~Sonner, ``{Matrix
  Models for Eigenstate Thermalization},''
  \href{http://dx.doi.org/10.1103/PhysRevX.13.031033}{{\em Phys. Rev. X}
  {\bfseries 13} no.~3, (2023) 031033},
  \href{http://arxiv.org/abs/2209.02130}{{\ttfamily arXiv:2209.02130
  [hep-th]}}.

\bibitem{Jafferis:2022wez}
D.~L. Jafferis, D.~K. Kolchmeyer, B.~Mukhametzhanov, and J.~Sonner,
  ``{Jackiw-Teitelboim gravity with matter, generalized eigenstate
  thermalization hypothesis, and random matrices},''
  \href{http://dx.doi.org/10.1103/PhysRevD.108.066015}{{\em Phys. Rev. D}
  {\bfseries 108} no.~6, (2023) 066015},
  \href{http://arxiv.org/abs/2209.02131}{{\ttfamily arXiv:2209.02131
  [hep-th]}}.

\bibitem{DiUbaldo:2023qli}
G.~Di~Ubaldo and E.~Perlmutter, ``{AdS$_{3}$/RMT$_{2}$ duality},''
  \href{http://dx.doi.org/10.1007/JHEP12(2023)179}{{\em JHEP} {\bfseries 12}
  (2023) 179}, \href{http://arxiv.org/abs/2307.03707}{{\ttfamily
  arXiv:2307.03707 [hep-th]}}.

\bibitem{Sonner:2017hxc}
J.~Sonner and M.~Vielma, ``{Eigenstate thermalization in the Sachdev-Ye-Kitaev
  model},'' \href{http://dx.doi.org/10.1007/JHEP11(2017)149}{{\em JHEP}
  {\bfseries 11} (2017) 149}, \href{http://arxiv.org/abs/1707.08013}{{\ttfamily
  arXiv:1707.08013 [hep-th]}}.

\bibitem{Pollack:2020gfa}
J.~Pollack, M.~Rozali, J.~Sully, and D.~Wakeham, ``{Eigenstate Thermalization
  and Disorder Averaging in Gravity},''
  \href{http://dx.doi.org/10.1103/PhysRevLett.125.021601}{{\em Phys. Rev.
  Lett.} {\bfseries 125} no.~2, (2020) 021601},
  \href{http://arxiv.org/abs/2002.02971}{{\ttfamily arXiv:2002.02971
  [hep-th]}}.

\bibitem{Belin:2020hea}
A.~Belin and J.~de~Boer, ``{Random statistics of OPE coefficients and Euclidean
  wormholes},'' \href{http://dx.doi.org/10.1088/1361-6382/ac1082}{{\em Class.
  Quant. Grav.} {\bfseries 38} no.~16, (2021) 164001},
  \href{http://arxiv.org/abs/2006.05499}{{\ttfamily arXiv:2006.05499
  [hep-th]}}.

\bibitem{Altland:2021rqn}
A.~Altland, D.~Bagrets, P.~Nayak, J.~Sonner, and M.~Vielma, ``{From operator
  statistics to wormholes},''
  \href{http://dx.doi.org/10.1103/PhysRevResearch.3.033259}{{\em Phys. Rev.
  Res.} {\bfseries 3} no.~3, (2021) 033259},
  \href{http://arxiv.org/abs/2105.12129}{{\ttfamily arXiv:2105.12129
  [hep-th]}}.

\bibitem{Chandra:2022bqq}
J.~Chandra, S.~Collier, T.~Hartman, and A.~Maloney, ``{Semiclassical 3D gravity
  as an average of large-c CFTs},''
  \href{http://dx.doi.org/10.1007/JHEP12(2022)069}{{\em JHEP} {\bfseries 12}
  (2022) 069}, \href{http://arxiv.org/abs/2203.06511}{{\ttfamily
  arXiv:2203.06511 [hep-th]}}.

\bibitem{Belin:2023efa}
A.~Belin, J.~de~Boer, D.~L. Jafferis, P.~Nayak, and J.~Sonner, ``{Approximate
  CFTs and Random Tensor Models},''
  \href{http://arxiv.org/abs/2308.03829}{{\ttfamily arXiv:2308.03829
  [hep-th]}}.

\bibitem{Mark_Srednicki_1999}
M.~Srednicki, ``The approach to thermal equilibrium in quantized chaotic
  systems,'' \href{http://dx.doi.org/10.1088/0305-4470/32/7/007}{{\em Journal
  of Physics A: Mathematical and General} {\bfseries 32} no.~7, (1999) 1163}.

\bibitem{Einstein_Brownian}
A.~Einstein, ``Über die von der molekularkinetischen theorie der wärme
  geforderte bewegung von in ruhenden flüssigkeiten suspendierten teilchen,''
  \href{http://dx.doi.org/https://doi.org/10.1002/andp.19053220806}{{\em
  Annalen der Physik} {\bfseries 322} no.~8, (1905) 549--560}.

\bibitem{Langevin_1908}
P.~Langevin, ``Sur la théorie du mouvement brownien,'' {\em C. R. Acad. Sci.
  Paris.} {\bfseries 146} (1908) 530–533.

\bibitem{Perrin_1909}
J.~Perrin, ``Mouvement brownien et réalité moléculaire,'' {\em Annales de
  chimie et de Physique} {\bfseries 18} no.~8, (1909) 5--114.

\bibitem{Ornstein_Uhlenbeck_1930}
G.~E. Uhlenbeck and L.~S. Ornstein, ``On the theory of the brownian motion,''
  \href{http://dx.doi.org/10.1103/PhysRev.36.823}{{\em Phys. Rev.} {\bfseries
  36} (1930) 823--841}.

\bibitem{Gardiner_Stochastic_Methods}
C.~Gardiner, {\em {Stochastic Methods: A Handbook for the Natural and Social
  Sciences}}.
\newblock Springer Series in Synergetics. Springer Berlin, Heidelberg, 2009.

\bibitem{Lennart_Lecture_notes}
L.~Sjögren, ``Lecture notes on stochastic processes,''.
  \url{http://gu-statphys.org/media/mydocs/LennartSjogren/kap6.pdf}.

\bibitem{Stanford:2020wkf}
D.~Stanford, ``{More quantum noise from wormholes},''
  \href{http://arxiv.org/abs/2008.08570}{{\ttfamily arXiv:2008.08570
  [hep-th]}}.

\bibitem{foini2019eigenstate}
L.~Foini and J.~Kurchan, ``Eigenstate thermalization hypothesis and out of time
  order correlators,'' \href{http://dx.doi.org/10.1103/PhysRevE.99.042139}{{\em
  Phys. Rev. E} {\bfseries 99} (2019) 042139}.

\bibitem{Foini-Pappalardi-Kurchan_free_prob}
S.~Pappalardi, L.~Foini, and J.~Kurchan, ``Eigenstate thermalization hypothesis
  and free probability,''
  \href{http://dx.doi.org/10.1103/PhysRevLett.129.170603}{{\em Phys. Rev.
  Lett.} {\bfseries 129} (2022) 170603}.

\bibitem{Henningson:1998gx}
M.~Henningson and K.~Skenderis, ``{The Holographic Weyl anomaly},''
  \href{http://dx.doi.org/10.1088/1126-6708/1998/07/023}{{\em JHEP} {\bfseries
  07} (1998) 023}, \href{http://arxiv.org/abs/hep-th/9806087}{{\ttfamily
  arXiv:hep-th/9806087}}.

\bibitem{Maldacena:2001kr}
J.~M. Maldacena, ``{Eternal black holes in anti-de Sitter},''
  \href{http://dx.doi.org/10.1088/1126-6708/2003/04/021}{{\em JHEP} {\bfseries
  04} (2003) 021}, \href{http://arxiv.org/abs/hep-th/0106112}{{\ttfamily
  arXiv:hep-th/0106112}}.

\bibitem{Collier:2023fwi}
S.~Collier, L.~Eberhardt, and M.~Zhang, ``{Solving 3d gravity with Virasoro
  TQFT},'' \href{http://dx.doi.org/10.21468/SciPostPhys.15.4.151}{{\em SciPost
  Phys.} {\bfseries 15} no.~4, (2023) 151},
  \href{http://arxiv.org/abs/2304.13650}{{\ttfamily arXiv:2304.13650
  [hep-th]}}.

\bibitem{Collier:2024mgv}
S.~Collier, L.~Eberhardt, and M.~Zhang, ``{3d gravity from Virasoro TQFT:
  Holography, wormholes and knots},''
  \href{http://arxiv.org/abs/2401.13900}{{\ttfamily arXiv:2401.13900
  [hep-th]}}.

\bibitem{Maldacena:2004rf}
J.~M. Maldacena and L.~Maoz, ``{Wormholes in AdS},''
  \href{http://dx.doi.org/10.1088/1126-6708/2004/02/053}{{\em JHEP} {\bfseries
  02} (2004) 053}, \href{http://arxiv.org/abs/hep-th/0401024}{{\ttfamily
  arXiv:hep-th/0401024}}.

\bibitem{Anous:2022wqh}
T.~Anous, M.~Meineri, P.~Pelliconi, and J.~d, ``{Sailing past the End of the
  World and discovering the Island},''
  \href{http://dx.doi.org/10.21468/SciPostPhys.13.3.075}{{\em SciPost Phys.}
  {\bfseries 13} no.~3, (2022) 075},
  \href{http://arxiv.org/abs/2202.11718}{{\ttfamily arXiv:2202.11718
  [hep-th]}}.

\bibitem{Maxfield:2016mwh}
H.~Maxfield, S.~Ross, and B.~Way, ``{Holographic partition functions and phases
  for higher genus Riemann surfaces},''
  \href{http://dx.doi.org/10.1088/0264-9381/33/12/125018}{{\em Class. Quant.
  Grav.} {\bfseries 33} no.~12, (2016) 125018},
  \href{http://arxiv.org/abs/1601.00980}{{\ttfamily arXiv:1601.00980
  [hep-th]}}.

\bibitem{deBoer:2024kat}
J.~de~Boer, D.~Liska, and B.~Post, ``{Multiboundary wormholes and OPE
  statistics},'' \href{http://arxiv.org/abs/2405.13111}{{\ttfamily
  arXiv:2405.13111 [hep-th]}}.

\bibitem{Hartman:2014oaa}
T.~Hartman, C.~A. Keller, and B.~Stoica, ``{Universal Spectrum of 2d Conformal
  Field Theory in the Large c Limit},''
  \href{http://dx.doi.org/10.1007/JHEP09(2014)118}{{\em JHEP} {\bfseries 09}
  (2014) 118}, \href{http://arxiv.org/abs/1405.5137}{{\ttfamily arXiv:1405.5137
  [hep-th]}}.

\bibitem{Collier:2019weq}
S.~Collier, A.~Maloney, H.~Maxfield, and I.~Tsiares, ``{Universal dynamics of
  heavy operators in CFT$_{2}$},''
  \href{http://dx.doi.org/10.1007/JHEP07(2020)074}{{\em JHEP} {\bfseries 07}
  (2020) 074}, \href{http://arxiv.org/abs/1912.00222}{{\ttfamily
  arXiv:1912.00222 [hep-th]}}.

\bibitem{DORN1994375}
H.~Dorn and H.-J. Otto, ``Two- and three-point functions in liouville theory,''
  \href{http://dx.doi.org/https://doi.org/10.1016/0550-3213(94)00352-1}{{\em
  Nuclear Physics B} {\bfseries 429} no.~2, (1994) 375--388}.

\bibitem{ZAMOLODCHIKOV1996577}
A.~Zamolodchikov and A.~Zamolodchikov, ``Conformal bootstrap in liouville field
  theory,''
  \href{http://dx.doi.org/https://doi.org/10.1016/0550-3213(96)00351-3}{{\em
  Nuclear Physics B} {\bfseries 477} no.~2, (1996) 577--605}.

\bibitem{Cho:2017oxl}
M.~Cho, S.~Collier, and X.~Yin, ``{Recursive Representations of Arbitrary
  Virasoro Conformal Blocks},''
  \href{http://dx.doi.org/10.1007/JHEP04(2019)018}{{\em JHEP} {\bfseries 04}
  (2019) 018}, \href{http://arxiv.org/abs/1703.09805}{{\ttfamily
  arXiv:1703.09805 [hep-th]}}.

\bibitem{Yellow_Book}
P.~di~Francesco, P.~Mathieu, and D.~Sénéchal,
  \href{http://dx.doi.org/https://doi.org/10.1007/978-1-4612-2256-9}{{\em
  {Conformal Field Theory}}}.
\newblock Springer New York, NY, 1996.

\bibitem{Fitzpatrick:2015zha}
A.~L. Fitzpatrick, J.~Kaplan, and M.~T. Walters, ``{Virasoro Conformal Blocks
  and Thermality from Classical Background Fields},''
  \href{http://dx.doi.org/10.1007/JHEP11(2015)200}{{\em JHEP} {\bfseries 11}
  (2015) 200}, \href{http://arxiv.org/abs/1501.05315}{{\ttfamily
  arXiv:1501.05315 [hep-th]}}.

\bibitem{Anous:2019yku}
T.~Anous and J.~Sonner, ``{Phases of scrambling in eigenstates},''
  \href{http://dx.doi.org/10.21468/SciPostPhys.7.1.003}{{\em SciPost Phys.}
  {\bfseries 7} (2019) 003}, \href{http://arxiv.org/abs/1903.03143}{{\ttfamily
  arXiv:1903.03143 [hep-th]}}.

\bibitem{Gerbershagen:2021yma}
M.~Gerbershagen, ``{Monodromy methods for torus conformal blocks and
  entanglement entropy at large central charge},''
  \href{http://dx.doi.org/10.1007/JHEP08(2021)143}{{\em JHEP} {\bfseries 08}
  (2021) 143}, \href{http://arxiv.org/abs/2101.11642}{{\ttfamily
  arXiv:2101.11642 [hep-th]}}.

\bibitem{Bouverot-Dupuis:2024nxk}
O.~Bouverot-Dupuis, S.~Pappalardi, J.~Kurchan, A.~Polkovnikov, and L.~Foini,
  ``{Random matrix universality in dynamical correlation functions at late
  times},'' \href{http://arxiv.org/abs/2407.12103}{{\ttfamily arXiv:2407.12103
  [cond-mat.stat-mech]}}.

\bibitem{Almheiri:2019hni}
A.~Almheiri, R.~Mahajan, J.~Maldacena, and Y.~Zhao, ``{The Page curve of
  Hawking radiation from semiclassical geometry},''
  \href{http://dx.doi.org/10.1007/JHEP03(2020)149}{{\em JHEP} {\bfseries 03}
  (2020) 149}, \href{http://arxiv.org/abs/1908.10996}{{\ttfamily
  arXiv:1908.10996 [hep-th]}}.

\bibitem{Lin:2020mme}
H.~W. Lin, ``{Bootstraps to strings: solving random matrix models with
  positivity},'' \href{http://dx.doi.org/10.1007/JHEP06(2020)090}{{\em JHEP}
  {\bfseries 06} (2020) 090}, \href{http://arxiv.org/abs/2002.08387}{{\ttfamily
  arXiv:2002.08387 [hep-th]}}.

\bibitem{Saad:2021rcu}
P.~Saad, S.~H. Shenker, D.~Stanford, and S.~Yao, ``{Wormholes without
  averaging},'' \href{http://arxiv.org/abs/2103.16754}{{\ttfamily
  arXiv:2103.16754 [hep-th]}}.

\bibitem{Karhunen47}
K.~Karhunen, ``{Uber lineare methoden in der wahrscheinlichkeitsrechnung},''
  {\em Annales Academiae Scientarum Fennicae} {\bfseries 37} (1947) 3--79.

\bibitem{Loeve48}
M.~Loève, {\em {Fonctions aleatoires du second ordre}}.
\newblock Processus Stochastiques et Mouve- ment Brownien, P. Levy (ed.), 1948.

\bibitem{deBoer:2008gu}
J.~de~Boer, V.~E. Hubeny, M.~Rangamani, and M.~Shigemori, ``{Brownian motion in
  AdS/CFT},'' \href{http://dx.doi.org/10.1088/1126-6708/2009/07/094}{{\em JHEP}
  {\bfseries 07} (2009) 094}, \href{http://arxiv.org/abs/0812.5112}{{\ttfamily
  arXiv:0812.5112 [hep-th]}}.

\bibitem{Son:2009vu}
D.~T. Son and D.~Teaney, ``{Thermal Noise and Stochastic Strings in AdS/CFT},''
  \href{http://dx.doi.org/10.1088/1126-6708/2009/07/021}{{\em JHEP} {\bfseries
  07} (2009) 021}, \href{http://arxiv.org/abs/0901.2338}{{\ttfamily
  arXiv:0901.2338 [hep-th]}}.

\bibitem{Sonner:2013mba}
J.~Sonner, ``{Holographic Schwinger Effect and the Geometry of Entanglement},''
  \href{http://dx.doi.org/10.1103/PhysRevLett.111.211603}{{\em Phys. Rev.
  Lett.} {\bfseries 111} no.~21, (2013) 211603},
  \href{http://arxiv.org/abs/1307.6850}{{\ttfamily arXiv:1307.6850 [hep-th]}}.

\bibitem{Jensen:2013ora}
K.~Jensen and A.~Karch, ``{Holographic Dual of an Einstein-Podolsky-Rosen Pair
  has a Wormhole},''
  \href{http://dx.doi.org/10.1103/PhysRevLett.111.211602}{{\em Phys. Rev.
  Lett.} {\bfseries 111} no.~21, (2013) 211602},
  \href{http://arxiv.org/abs/1307.1132}{{\ttfamily arXiv:1307.1132 [hep-th]}}.

\bibitem{Post:2022dfi}
B.~Post, J.~van~der Heijden, and E.~Verlinde, ``{A universe field theory for JT
  gravity},'' \href{http://dx.doi.org/10.1007/JHEP05(2022)118}{{\em JHEP}
  {\bfseries 05} (2022) 118}, \href{http://arxiv.org/abs/2201.08859}{{\ttfamily
  arXiv:2201.08859 [hep-th]}}.

\bibitem{Altland:2022xqx}
A.~Altland, B.~Post, J.~Sonner, J.~van~der Heijden, and E.~P. Verlinde,
  ``{Quantum chaos in 2D gravity},''
  \href{http://dx.doi.org/10.21468/SciPostPhys.15.2.064}{{\em SciPost Phys.}
  {\bfseries 15} no.~2, (2023) 064},
  \href{http://arxiv.org/abs/2204.07583}{{\ttfamily arXiv:2204.07583
  [hep-th]}}.

\bibitem{Pelliconi:2023ojb}
P.~Pelliconi and J.~Sonner, ``{The influence functional in open holography:
  entanglement and R\'enyi entropies},''
  \href{http://dx.doi.org/10.1007/JHEP06(2024)185}{{\em JHEP} {\bfseries 06}
  (2024) 185}, \href{http://arxiv.org/abs/2310.13047}{{\ttfamily
  arXiv:2310.13047 [hep-th]}}.

\bibitem{Hernandez-Cuenca:2024pey}
S.~Hern\'andez-Cuenca, ``{Wormholes and Factorization in Exact Effective
  Theory},'' \href{http://arxiv.org/abs/2404.10035}{{\ttfamily arXiv:2404.10035
  [hep-th]}}.

\end{thebibliography}\endgroup

\appendix

\section{Sum over images} \label{app:Sum_over_images}

The full answer for the thermal correlator is \cite{Maldacena:2001kr} 
\begin{equation}
    G_{\beta}(t) = \left(\frac{2 \pi}{\beta} \right)^{2 \Delta} \sum_{n \in \mathbb Z} \frac{1}{\left[ 2 \cosh(\frac{2 \pi t}{\beta}) + 2 \cosh(\frac{4 \pi^2 n}{\beta}) \right]^{\Delta}}
\end{equation}
The contribution for $n = 0$ is the leading one reported in~\eqref{avg_two_pt_gravity}, while all the others represent geodesics that wrap around the black hole $n$-times. In the limit considered in the text, $\beta, \Delta \to 0$ with $2 \pi \Delta/\beta \equiv \gamma$ fixed, these contribution can be resummed. In particular, in this limit it is possible to show that
\begin{equation}
    \lim_{\Delta, \beta \to 0} \, \left[ \cosh(\frac{2 \pi t}{\beta}) + \cosh(\frac{4 \pi^2 n}{\beta}) \right]^{\Delta} \, = \; \begin{cases}
e^{2 \pi \gamma |n|} \quad &|t| < 2 \pi |n| \ ,\\
e^{ \gamma |t|} \quad &|t| \geq 2 \pi |n| \ .
\end{cases}
\end{equation}
Therefore, it is not hard to re-sum the whole series. In particular, we have
\begin{equation}
   \lim_{\Delta, \beta \to 0} \, G_{\beta}(t) = e^{- \gamma |t|} + 2 \, |n| \, e^{-\gamma |t|} + \frac{2}{e^{2 \pi \gamma} - 1} \, e^{- 2 \pi \gamma |n|} \ , \qquad 2 \pi n \leq t < 2 \pi (n + 1) \ ,
\end{equation}
where we remind the reader that $n$ is an integer. This is the exact result, which shows the correction to the exponential behavior presented in the main text. We can approximate this piece-wise function with a smooth one setting $t \approx 2 \pi n$. This approximation becomes exact in the limit $2 \pi \gamma \to 0$, but it is good generically. Thus, taking into account all the images, the correlation function~\eqref{avg_two_pt_gravity} becomes
\begin{equation}
   \lim_{\Delta, \beta \to 0} \, G_{\beta}(t) \approx  \left( \coth(\pi \gamma) +  \frac{|t|}{\pi} \right)\, e^{-\gamma |t|}  \ .
\end{equation}

\section{Autocorrelation function in the CFT} \label{sec:autocorrelation_in_CFT}

In order to compute the autocorrelation function, we start from the necklace expansion
\begin{equation}
    \mathbb E[G_{\beta} (z, \bar z) G_{\beta} (w, \bar w)] = \frac{1}{Z^2(\tau, \bar \tau)} \sum_{p, q} \mathbb E \Big[|C_{\cO p_1 p_2}|^2 |C_{\cO q_1 q_2}|^2 \Big]  \, \Big| \cF_{\rm N} (p_1,p_2| \tau, z)  \cF_{\rm N} ( q_1, q_2|  \tau,  w) \Big|^2 \ .
\end{equation}
In order to compute the expectation value, we have to consider all the Wick contractions between the indices. One is clearly the disconnected one, but as in Section~\ref{BM_and_thermal_two_pt} we have two additional ones, where we set $p_1$ equal to $q_1$ and $p_2$ equal to $q_2$, and viceversa. This in turn gives
\begin{multline}
    \mathbb E[G_{\beta} (z, \bar z) G_{\beta} (w, \bar w)] \, = \,  \mathbb E[G_{\beta} (z, \bar z)] \, \mathbb E[G_{\beta} (w, \bar w)] \, + \\
    \frac{1}{Z^2(\tau, \bar \tau)} \, \Bigg[ \; \bigg| \int_{\frac{c-1}{24}}^{\infty} \de p_1\, \de p_2 \, \rho(p_1) \, \rho(p_2) \, C_0^2(\cO, p_1, p_2) \, \cF_{\rm N} (p_1, p_2| \tau, z) \, \cF_{\rm N} (p_2, p_1| \tau, w) \bigg|^2 \\
    + \; \bigg| \int_{\frac{c-1}{24}}^{\infty} \de p_1\, \de p_2 \, \rho(p_1) \, \rho(p_2) \, C_0^2(\cO, p_1, p_2) \, \cF_{\rm N} (p_1, p_2| \tau,  z) \, \cF_{\rm N} (p_1, p_2| \tau, w) \bigg|^2 \; \Bigg] \ .
\end{multline}
Using standard manipulations (see \cite{Chandra:2022bqq} for details), we can rewrite it as
\begin{multline}
    \mathbb E[G_{\beta} (z, \bar z) G_{\beta} (w, \bar w)] \, = \,  \mathbb E[G_{\beta} (z, \bar z)] \, \mathbb E[G_{\beta} (w, \bar w)] \, + \\
    \frac{1}{Z^2(\tau, \bar \tau)} \, \Bigg[ \; \bigg| \int_{\frac{c-1}{24}}^{\infty} \de p_1\, \de p_2 \, \rho(p_1) \, \rho(p_2) \, C_0^2(\cO, p_1, p_2) \, \cF_{\rm N} (p_1, p_2| \tau, z) \, \bar \cF_{\rm N} (p_1, p_2| - \tau, - w) \bigg|^2 \\
    + \; \bigg| \int_{\frac{c-1}{24}}^{\infty} \de p_1\, \de p_2 \, \rho(p_1) \, \rho(p_2) \, C_0^2(\cO, p_1, p_2) \, \cF_{\rm N} (p_1, p_2| \tau,  z) \, \bar \cF_{\rm N} (p_1, p_2| -\tau, w) \bigg|^2 \; \Bigg] \ .
\end{multline}
This agrees with the result \eqref{two_pt_fn_thermal} reported in the main text.

\end{spacing}
\end{document}